\documentclass[12pt]{revtex4} 
 \usepackage{graphicx}

 \def\be{\begin{equation}}
 \def\ee{\end{equation}}
 \def\bea{\begin{eqnarray}}
 \def\eea{\end{eqnarray}}
 \def\bean{\begin{eqnarray*}}
 \def\eean{\end{eqnarray*}}
 \def\D{\displaystyle}
 \def\T{\textstyle}
 \def\p{\partial}
 \def\l{\left}
 \def\r{\right}
 \def\Im{{\rm Im}}
 \def\Re{{\rm Re}}
 \def\bm#1{\mbox{\boldmath$#1$}}
 \def\SumInt{\hbox{$\sum$}\hspace{-1.1em}\int\,}
 \def\ksim{\mathrel{\rlap{\lower0.2em\hbox{$\sim$}}\raise0.2em\hbox{$<$}}}
 \def\gsim{\mathrel{\rlap{\lower0.2em\hbox{$\sim$}}\raise0.2em\hbox{$>$}}}
\begin{document}

\title{Hard gluon damping in hot QCD}
\author{A.~Peshier}
\email{Andre.Peshier@theo.physik.uni-giessen.de}             
\affiliation{Institut f{\"u}r Theoretische Physik, 
             Universit{\"a}t Giessen,
             35392 Giessen, Germany}

\begin{abstract}
The gluon collisional width in hot QCD plasmas is discussed with emphasis on temperatures near $T_c$, where the coupling is large.
Considering its effect on the entropy, which is known from lattice calculations, it is argued that the width, which in the perturbative limit is given by $\gamma \sim g^2 \ln(1/g)\, T$, should be sizeable at intermediate temperatures but has to be small close to $T_c$.
Implications of these results for several phenomenologically relevant quantities, such as the energy loss of hard jets, are pointed out.
\end{abstract}

\maketitle

\section{Introduction}
The dispersion relation and the damping rate of single-particle excitations in many-particle systems are closely related to a variety of phenomenologically important quantities.
In a QCD plasma at temperatures much higher than the transition temperature $T_c \sim 200\,$MeV, where the coupling $g$ is small and perturbation theory should be applicable, the quark and gluon excitation energies follow directly from the real part of the 1-loop self-energies, which are of the order $(gT)^2$.
The calculation of their width, however, requires already to lowest order in $g$ a summation of infinitely many diagrams. Resumming hard thermal loops (HTL), Braaten and Pisarski \cite{BraatP} obtained the widths of quarks and gluons at rest, which are proportional to $g^2 T$.
The case of excitations with a finite momentum is more intricate because their width diverges even after the HTL resummation due to the exchange of soft magnetic gluons. With a cutoff of the order of $g^2 T$, either a magnetic mass and/or the width itself, the width of moving charged excitations is $\sim g^2 \ln(1/g)\, T$ on rather general grounds \cite{Pisar89}. Even in the case of weak coupling, the gluon width tests directly the nonperturbative sector of QCD.

A similar breakdown of perturbation theory occurs also in the calculation of the thermodynamic potential $\Omega(T)=-p(T)V$, at order ${\cal O}(g^6)$ \cite{Linde}.
The expansion in $g$, which is known up to ${\cal O}(g^6 \ln g^{-1})$ \cite{KajanLRS}, is not reliable in the physically interesting regime as  probed in present relativistic heavy ion experiments. For the large coupling strength expected at temperatures near $T_c$, it does not converge but shows a behavior typical of asymptotic series. 
In fact, one can hardly expect a converging expansion since it has to be defined in a circle in the complex plane, while in QCD a transition $g^2 \rightarrow -g^2$ is presumably non-analytic. In perturbation theory, this manifests itself in the number of diagrams increasing rapidly with the order.
A strategy to remedy the situation in practical calculations is a partial resummation of the perturbative series, taking into account those classes of diagrams whose number increase most rapidly. These are related to the various orders in the loop expansion in the $\Phi$-derivable approximation scheme \cite{Baym} to be utilized below. In this scheme, the thermodynamic potential is expressed in terms of {\em dressed} propagators, which are determined self-consistently.
The restriction to the leading loop order(s) for large coupling may seem counter-intuitive. Since it is essential for the following, it is worth mentioning another, yet related, motivation: An asymptotic series, to give the best approximation possible, should be truncated at an order related {\em inversely} to the coupling; for the QCD thermodynamic potential near $T_c$ possibly already at the order ${\cal O}(g^2)$ \cite{Peshi02}. At large coupling, such a perturbative result is, however, not thermodynamically consistent since various thermodynamic quantities are related to each other by derivatives with respect to the temperature. Since $T$ is also the relevant scale in the running coupling, which is large, a thermodynamically consistent approximation (when expanded in $g$) has to contain {\em some} contributions of higher orders.

A truncation of a resummation scheme based on 2-point functions is, a priory, delicate for QCD because of gauge invariance.
This problem can be evaded by receding to approximately self-consistent resummations of the thermodynamic potential using relevant gauge-invariant contributions to the propagators. Indeed, results calculated with HTL propagators \cite{BlaizIR, Peshi} agree with QCD lattice data down to temperatures of about $3T_c$ \footnote{
    Note that the HTL propagators, while derived for soft momenta,
    have the correct limit for the thermodynamically relevant large
    momenta.
    Similar results have also been obtained within HTL perturbation
    theory \cite{AnderBPS}.}.
The HTL propagators can be reduced even further by neglecting the Landau-damping parts and retaining only the dominant pole contributions, approximating as well the dispersion relations by the asymptotic mass shells. The resulting phenomenological models \cite{pQP} can describe the lattice data even close to $T_c$ because they allow for an IR-enhancement of the running coupling compared to the 2-loop formula used in the HTL calculations \footnote{
    For another quasiparticle model see \cite{SchneW}.}.
In all of these approaches the observed decrease in the effective degrees of freedom near $T_c$ is directly related to the temperature dependent mass scale $m \sim gT$ that characterizes the excitations. Interpreted as quasiparticles, they become heavy near $T_c$ due to the running coupling.
While apparently this reflects important interaction effects (as motivated above), so far none of the approaches takes into consideration the expected width of the quasiparticles. This, however, is a priori not justified for large coupling, when the width might become comparable to the mass of the quasiparticles \cite{JuchCG}.

In principle, the dressed propagators and the widths can be calculated, by Schwinger-Dyson equations, in the $\Phi$-derivable approximation scheme. However, apart from the aforementioned sensitivity of the width to the soft QCD sector, there is the basic requirement of gauge invariance of physical quantities such as the width itself or the deduced pressure. Moreover, 
the resummed propagators need to be renormalized nonperturbatively.
Notwithstanding the recent progress in these issues \cite{ArrizS,vHeesK,BlaizIReinosa}, the problem is involved.
It therefore seems interesting to ask a reversed and simpler question: What can be inferred about the propagators from other quantities which can be reliably calculated by other means? An evident choice for such a quantity is the thermodynamic potential, which can be numerically calculated rather precisely on the lattice. 

Clearly, one cannot expect detailed information from the bulk properties of the many-particle system, but estimates for some important features seem feasible. Indeed, the large quasiparticle masses `predicted' in \cite{BlaizIR,Peshi,pQP} compare well with direct results from lattice QCD \cite{Petre}.
It is obvious from the available phase space, though, that only properties of excitations with hard momenta, $k \gsim T$, can be accessible.
In the following I focus on the entropy, $s = -\p\Omega/\p T$. Since it provides a measure of the population of phase space, one expects an increased entropy for a system of off-shell particles as described by the width.
From the estimate $\gamma \sim g^2 \ln(1/g)\, T$, the width might become large with increasing $g$, leading to the principal question whether this would be reconcilable with the small entropy near $T_c$, as calculated on the lattice.

To approach this question, in this paper some general relations between propagators and entropy will be discussed. Section \ref{sec 2} starts with a brief outline of the formalism of self-consistent approximations. In Section \ref{sec 3}, the case of particles with a Lorentz spectral function is considered in some detail, followed by an analysis of the sensitivity of the results on the spectral function. For the sake of transparency a scalar field theory is discussed before switching over to QCD in Section \ref{sec 4}. Given the remarkably universal scaling behavior of the QCD entropy for various numbers of quark flavors \cite{KarscLP}, I consider here the representative case of the quenched limit of QCD. In the conclusions, some implications of the present findings are pointed out. Some formal details were deferred to the Appendix.

\section{Propagator and thermodynamics \label{sec 2}}

Following the work of Luttinger and Ward \cite{LuttiW}, the thermodynamic potential $\Omega$ of a system of particles with a given interaction can be expressed in terms of the exact 2-point function(s).
Considering for simplicity a scalar theory with the propagator $\Delta$, the expression reads in the imaginary-time formalism (setting the volume $V=1$) \be
  \Omega
  =
  \T\frac12\,
  \D\SumInt\l( \ln(-\Delta^{\!-1}) + \Pi\Delta\rule{0em}{1.2em} \r)
  -\Phi[\Delta] \, .
  \label{eq: Omega LW}
\ee
The self-energy $\Pi = \Delta_0^{\!-1}-\Delta^{\!-1}$ is given by the sum of all 2-particle irreducible skeleton diagrams, which is closely related to the expansion of the functional $\Phi$. The right hand side of (\ref{eq: Omega LW}) is a functional to be evaluated with the exact propagator $\Delta$, which is obtained from the stationarity condition $\delta\Omega[\Delta]/\delta\Delta = 0$. The functional variation leads to
\be
  \Pi = 2\,\frac{\delta\Phi}{\delta\Delta} \, ,
  \label{eq: Pi_sc}
\ee
i.\,e., the self-energy is self-consistently obtained by cutting a full propagator line in the 2-particle irreducible skeleton expansion of $\Phi$.
For a $(\kappa/3!)\, \phi^3 + (g^2/4!)\, \phi^4$ interaction, the functional reads
\[
  \Phi
  =
  3\, \raisebox{-.6em}{\includegraphics[scale=0.7]{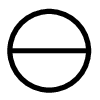}}
  +
  3\, \raisebox{-.6em}{\includegraphics[scale=0.7]{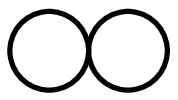}}
  +
  12\! \raisebox{-.8em}{\includegraphics[scale=0.5]{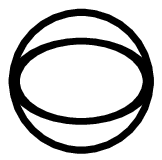}}\!
  + \,\ldots
\]
From this exact representation, symmetry-conserving (`$\Phi$-derivable') approximations \cite{Baym} follow by truncating the $\Phi$-expansion at a given loop order. For a given order, the self-energy is then calculated self-consistently according to Eq.~(\ref{eq: Pi_sc}).

To derive the entropy in terms of the resummed propagator, the Matsubara-sum in (\ref{eq: Omega LW}) is first transformed into a contour integral in the complex energy plane. After wrapping the contour around the real axis one obtains first
\be
  \Omega
  =
  \int_{k^4} n(\omega)\,
  \Im\!\l( \ln(-\Delta^{\!-1}) + \Pi\Delta \rule{0em}{1.2em} \r)
  -\Phi\, ,
\ee
where $\Delta$ now denotes the retarded propagator; $\int_{k^4} = \int_{k^3} \int d\omega/(2\pi)$, $\int_{k^3} = \int d^3k/(2\pi)^3$, and $n(\omega) = (\exp(\omega/T)-1)^{-1}$ is the Bose-Einstein distribution function.
Taking $\delta\Omega/\delta\Delta = 0$ into account leads to
\be
  s
  =
  -\frac{\p\Omega}{\p T}
  =
  -\int_{k^4} \frac{\p n}{\p T}\, \Im\!\l(
    \ln(-\Delta^{\!-1}) + \Pi\Delta\rule{0em}{1.2em}
   \r)
  +\l.\frac{\p\Phi}{\p T}\r|_{\Delta} .
\ee
This expression can be decomposed, $s = s^{dqp}+s'$, where
\be
  s^{dqp}
  =
  -\int_{k^4} \frac{\partial n}{\partial T}\,
   \l(
    \Im\ln(-\Delta^{-1}) + \Im\Pi\, \Re\Delta \rule{0em}{1.2em}
   \r) ,
  \label{eq: s dqp}
\ee
and
\be
  s'
  =
  -\int_{k^4} \frac{\partial n}{\partial T}\, \Re\Pi\, \Im\Delta
  +\l.\frac{\p\Phi}{\p T}\r|_{\Delta} .
\ee
Remarkably, in $s'$ the first term cancels the contributions from $\Phi$ with one and two vertices shown diagrammatically above, see e.\,g.\ \cite{BlaizIR}. In other words, the contribution (\ref{eq: s dqp}), given the propagator, is the leading-loop resummed entropy in the $\Phi$-derivable scheme. According to the arguments put forward in the Introduction, it is a preferable approximation of the exact entropy at large coupling \footnote{
    With regard to the application in QCD it is emphasized that the
    cancellation in $s'$ holds for any propagator, in particular for 
    the exact one. Parameterizing later the exact gluon propagator by the
    dispersion relation and the width, which are physical quantities,
    ensures the gauge invariance of the results.
    At the same time, reasoning along the same lines as in the 
    Introduction, the deviation of the entropy from the self-consistent 
    and from the exact propagator should be small.},
\be
  s \approx s^{dqp} \, .
  \label{eq: sdqp}
\ee
The cancellation in $s'$ is basically a topological feature, and 
expressions analogous to (\ref{eq: s dqp}) hold also in other theories
\cite{sFctnl,BlaizIR}, including QCD. In the context of the Fermi liquid theory it is called the {\em dynamical quasiparticle contribution} to the entropy \cite{CarneP}.
Although the approximation (\ref{eq: s dqp}) is a nonperturbative resummation, it has a simple 1-loop structure and it does not depend on the vertices.
The integrand can be rewritten using
\bean
  \Im\ln(-\Delta^{-1})
  &=&
  \pi\, {\rm sgn}(\Im\Delta) - {\rm arg}(\Delta)
  \\
  &=&
  \pi\, {\rm sgn}(\Im\Delta)\, \Theta(\Re\Delta)
  -{\rm arctan}(\Im\Delta/\Re\Delta) \, .
\eean
In the first term, the real part of the propagator is negative for small $\omega>0$, as shown below, and it changes sign at $\omega^2 = \omega_k^2$.
Using ${\rm sgn}(\Im\Delta(\omega)) = -{\rm sgn}(\omega)$, this term yields the expression of the entropy of free bosons with the dispersion relation $\omega_k$ (and zero width),
\be
  s^{(0)}
  =
  \frac1T
  \int_{k^3}
  \l( -T\ln\l(1-e^{-\omega_k/T}\r) + \omega_k\, n(\omega_k) \r) .
  \label{eq: s(0)}
\ee
Below, $\omega_k$ will be simplify referred to as the dispersion relation of the dynamical quasiparticles although it is not the real part of the pole of the propagator (unless its imaginary part vanishes).
In the total entropy, 
\be
  s = s^{(0)}+\Delta s \, ,
\ee
(where now the relation (\ref{eq: sdqp}) simplifies the notation) the effects of a non-zero spectral width are solely due to the second contribution,
\be
  \Delta s
  =
  \int_{k^4} \frac{dn}{dT}\,
  \l( {\rm arctan}\, \lambda - \frac\lambda{1+\lambda^2} \r) ,
  \label{eq: Delta s}
\ee
where $\lambda = \Im\Delta/\Re\Delta$.
For later reference it is noted that the second term in the parenthesis is $\Im\Delta$ times the derivative of the phase ${\rm arctan}\, \lambda$ with respect to $\Im\Delta$.

The expectation that the entropy is increased for a non-zero spectral width, $\Delta s > 0$, can be verified from the representation (\ref{eq: Delta s}) under rather general assumptions.
To this end, the analytic propagator, for complex $k_0$, is first expressed in the Lehmann representation,
\[
  \Delta^a(k_0,\bm k)
  =
  \int_{-\infty}^\infty \frac{d\omega}{2\pi}\,
    \frac{\rho(\omega,\bm k)}{k_0-\omega} \, .
\]
The spectral function is the discontinuity of the propagator at the real axis,
\be
  \rho(\omega, \bm{k})
  =
  \Delta^a(\omega-i\epsilon,\bm{k})
  - \Delta^a(\omega+i\epsilon,\bm{k}) \, .
\ee
It is real, odd in $\omega$ with $\omega\rho(\omega) \ge 0$, and it satisfies the sum rule
\be
  \int_{-\infty}^\infty \frac{d\omega}{2\pi}\, \omega \rho(\omega, \bm k)
  =
  1
  \label{eq: sum rule}
\ee
for all values of \bm{k}.
This implies that the propagator approaches the free limit at large $k_0$, \be
  \Delta^a(k_0,\bm k)
  =
  \int_0^\infty \frac{d\omega}\pi\,
    \omega\, \frac{\rho(\omega,\bm k)}{k_0^2-\omega^2} \,
  \stackrel{k_0 \rightarrow \infty}{\longrightarrow} \,
  \frac1{k_0^2} \, .
\ee
These general properties of the spectral function have several implications for the retarded propagator $\Delta(\omega) = \Delta^a(\omega+i\epsilon)$.
Its imaginary part, which by the reflection principle is $-\frac12\,\rho(\omega)$, satisfies
\bean
  && \Im\Delta(\omega=0) \, = \, 0 \, , \\
  && \Im\Delta \le 0 \quad \mbox{for \ } \omega>0, \\
  && \Im\Delta \rightarrow 0 \quad \mbox{for \ }
    \omega \rightarrow \infty.
\eean
Similarly, one readily infers
\bean
  && \Re\Delta(\omega=0) 
    = -\int\frac{d\omega'}{2\pi}\, 
       \frac{\rho(\omega')}{\omega'}\, < \, 0 \, ,  \\
  && \Re\Delta \rightarrow \omega^{-2}
   \quad \mbox{for \ } \omega \rightarrow \infty ,
\eean
and that odd-order derivatives of Re$\Delta(\omega)$ vanish at $\omega=0$.
Now consider a generic spectral function with a prominent peak near $\omega_k$ and a characteristic width $\gamma$, possibly with some additional minor substructures. In this case $\Re\Delta(\omega)$ changes its sign only once for $\omega>0$, i.\,e., the `dispersion relation' $\omega_k$ is unique, which will be the only assumption for the following argument.
In principle then, there are two typical cases of propagators, see Fig.~\ref{fig: types of prop}:
\begin{figure}[ht]
 \hskip -1em
 \includegraphics[scale=0.75]{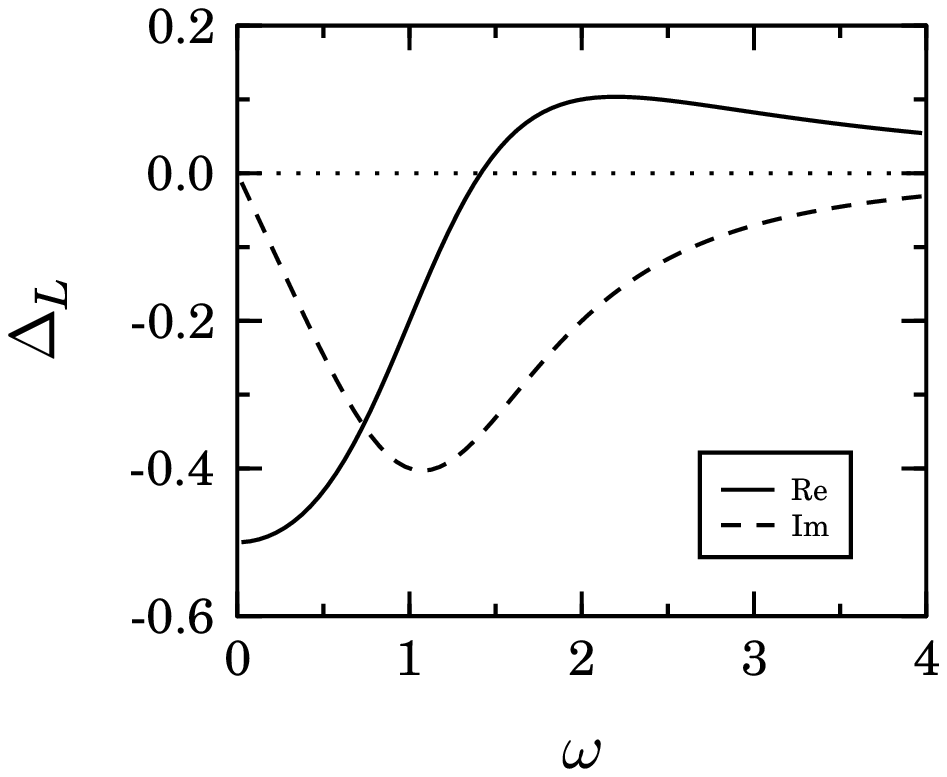}
 \hfill
 \includegraphics[scale=0.75]{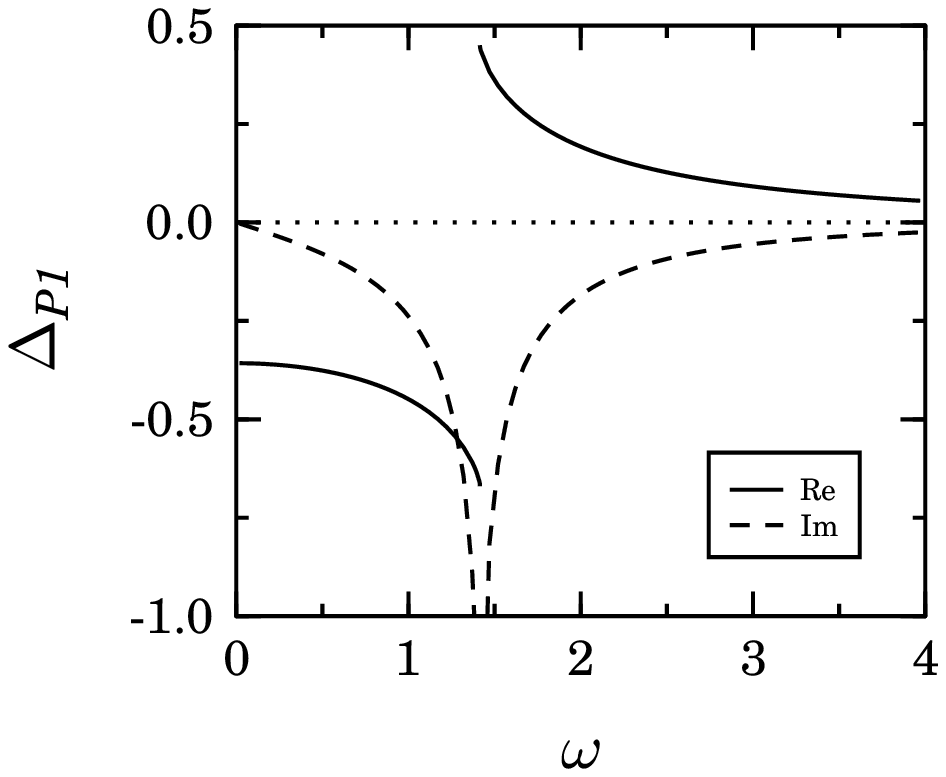}
 \caption{The real and the imaginary parts of the propagators $\Delta_L$
    and $\Delta_{P1}$ defined in Eq.~(\ref{eq: Delta_L}) and in 
    Sec.~\ref{sec: sensitivity on rho}, respectively. 
    All quantities are in units of the width $\gamma$, and the energy 
    scale $E$ in the spectral functions is chosen such that 
    $\omega_k=\sqrt2$.
  \label{fig: types of prop}}
\end{figure}
$(i)$ the imaginary part is regular, and the real part is smooth, and $(ii)$ the imaginary part is singular (but integrable due to (\ref{eq: sum rule})), and the real part is discontinuous.
Common to both cases is that $\omega_k$ is determined by the real part of the self-energy,
\[
  \mbox{Re}\,\Delta^{-1}(\omega_k)
  =
  \Delta_0^{-1}(\omega_k) - \mbox{Re}\,\Pi(\omega_k)
  =
  0 \, .
\]
For singular spectral functions the dispersion relation $\omega_k$ and the peak position in $\rho$ coincide, while they can be separated by an energy of the order of $\gamma$ for regular spectral functions.
The integrand of the entropy contribution $\Delta s$ is discontinuous at $\omega_k$; shown in Fig.~\ref{fig: intgrd_Delta_s} is the parenthesis term in Eq.~(\ref{eq: Delta s}).
\begin{figure}[ht]
 \hskip -1em
 \includegraphics[scale=0.75]{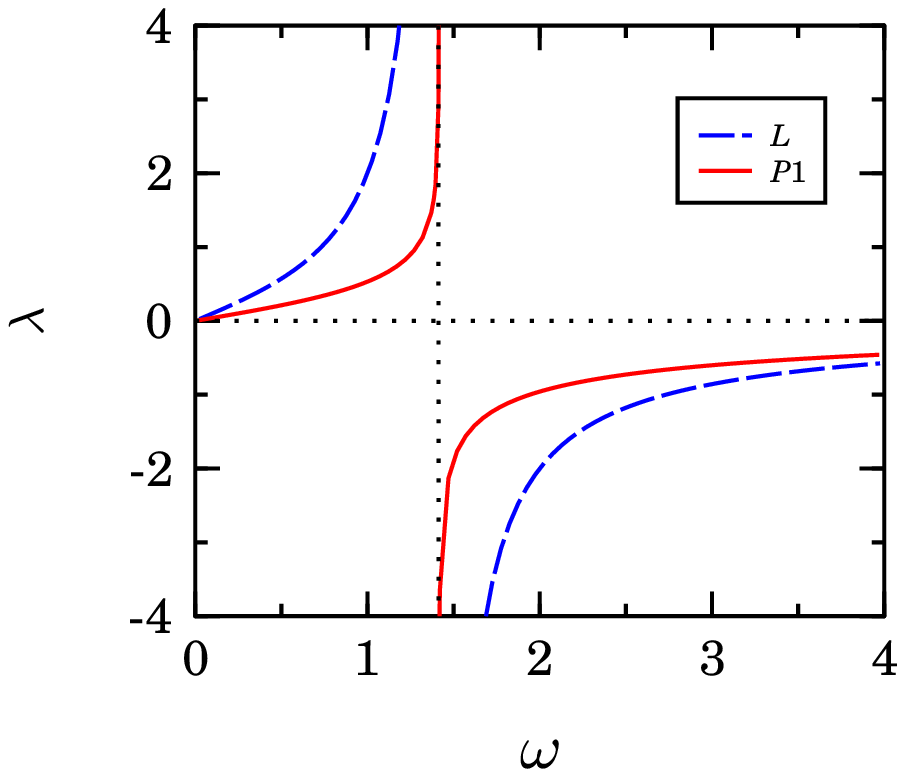}
 \hfill
 \includegraphics[scale=0.75]{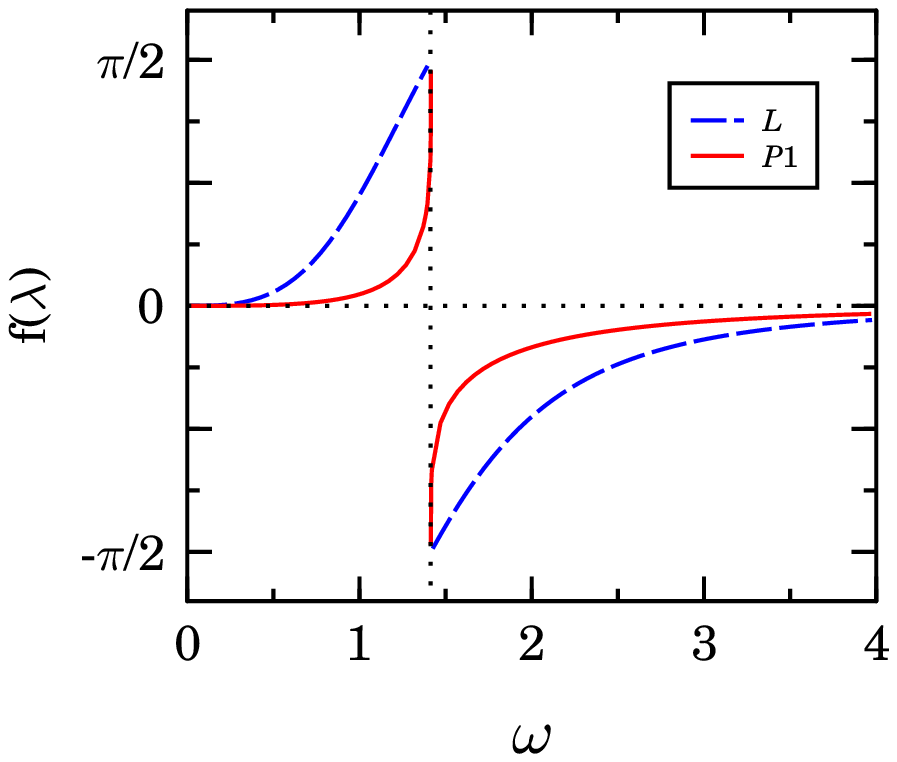}
 \caption{The functions $\lambda = \Im\Delta/\Re\Delta$, and 
    $f(\lambda) = \arctan\lambda-\lambda/(1+\lambda^2)$ (occurring in 
    Eq.~(\ref{eq: Delta s}) in the integrand of $\Delta s$) for the 
    propagators shown in Fig.~\ref{fig: types of prop}.
  \label{fig: intgrd_Delta_s}}
\end{figure}
This factor is, in an approximate way, symmetric near $\omega_k$, where $\lambda(\omega) = \Im\Delta/\Re\Delta$ is singular. Since the second factor in the integrand (\ref{eq: Delta s}),
\be
  \frac{dn}{dT} 
  = 
  \frac\omega{T^2}\, \frac{\exp(\omega/T)}{(\exp(\omega/T)-1)^2} \, ,
\ee  
is monotonically decreasing with $\omega$, it is plausible that $\Delta s > 0$. A rigorous argument is given in the Appendix.
From Fig.~\ref{fig: intgrd_Delta_s} it can also be expected that the entropy increase is smaller for singular spectral functions than for regular ones.

\section{Spectral functions \label{sec 3}}

\subsection{Lorentz spectral function}

In the previous section, the spectral function was introduced to deduce some general properties of the propagator and the entropy. At the same time, lacking a direct calculation, it is more intuitive (and more efficient due to the analytic properties) to model the spectral function rather than the propagator.
From the spectral function of free relativistic particles with $\Delta_0^{-1} = k_0^2-\omega_m^2$ where $\omega_m^2 = m^2+\bm k^2$, 
\[
  \rho_0(\omega) 
  = 
  2\pi \l[\delta\l( 
  (\omega - \omega_m)^2 \r) - \delta\l( (\omega + \omega_m)^2 \r)\r] \, ,
\]
an often used ansatz to describe non-zero width is obtained by replacing the $\delta$-function by a Lorentzian, 
\be
 \rho_L(\omega)
 =
 \frac\gamma{E} \l(
   \frac1{(\omega-E)^2+\gamma^2} - \frac1{(\omega+E)^2+\gamma^2}
 \r) .
 \label{eq: rhoL}
\ee
The corresponding retarded propagator can be easily calculated by a contour integration,
\[
  \Delta_L(\omega)
  =
  \frac1{\omega^2-E^2-\gamma^2+2i\gamma\omega} \, .
\]
In general, the analytical continuation of the retarded propagator to complex energies is analytic in the upper plane. In the present case it has poles in the lower plane, at $k_0 = \pm E - i\gamma$.
The parameter $E$ is directly related to the dispersion relation. 
Choosing $E^2(\bm{k}) = \bm k^2+m^2-\gamma^2$, the propagator becomes
\be
  \Delta_L(\omega, \bm k)
  =
  \frac1{\omega^2-\bm k^2-m^2+2i\gamma\omega} \, .
  \label{eq: Delta_L}
\ee
With this convention \footnote{
    There is no ambiguity in the spectral function (\ref{eq: rhoL}) for 
    $\gamma^2>\bm k^2+m^2$ as obvious from the alternative 
    representation $\rho_L = 4\gamma\omega / 
    (\omega^2-E^2-\gamma^2)^2+4\gamma^2\omega^2)$.     
    Note, however, that the spectral function becomes slightly more 
    asymmetric, and that the poles of the propagator (\ref{eq: Delta_L}), 
    at $k_0 = -i\gamma \pm (\omega_m^2 - \gamma^2)^{1/2}$, turn purely 
    imaginary for $\gamma > \omega_m$.
  \label{footnote1}}, 
the parameter $m^2$ corresponds directly to the real part of the retarded self-energy. This has the advantage that the dispersion relation does not depend on $\gamma$, $\omega_k = \omega_m$. 

Turning now to the entropy, one should note that in general the mass and the width parameters are momentum-dependent. The resulting effects will be considered below; for now the parameters are assumed to be constant.
For the propagator (\ref{eq: Delta_L}), the contribution (\ref{eq: s(0)}) to the dynamical quasiparticle entropy,
\be
  s_L^{(0)}(m)
  =
  \frac1T
  \int_{k^3} \l( -T\ln(1-e^{-\omega_m/T}) + \omega_m\, n(\omega_m/T) \r) ,
\ee
is simply the entropy of free bosons with mass $m$.
Corresponding expressions for QCD have been the starting point in the approaches \cite{pQP}, which interpreted the thermodynamically relevant transverse gluon and quark particle-excitations as quasiparticles with 
masses given by the asymptotic self-energies (and respective degeneracies).

The contribution (\ref{eq: Delta s}) due to the non-zero width reads explicitly
\be
  \Delta s_L(m,\gamma)
  =
  \int_{k^4}\frac{\partial n}{\partial T}\,
   \l(
   \arctan\frac{2\gamma\omega}{\omega_m^2-\omega^2}
   -2\gamma\omega
     \frac{\omega_m^2-\omega^2}
          {(\omega^2-\omega_m^2)^2+(2\gamma\omega)^2}
   \r).
\ee
A numerical integration shows -- in line with the general expectation -- that the total entropy $s_L=s_L^{(0)} + \Delta s_L$ increases with the width and decreases with $m$, cf.\ Fig.~\ref{fig: s(m,Gamma)}.
\begin{figure}[ht]
 \hskip -1em
 \includegraphics[scale=0.75]{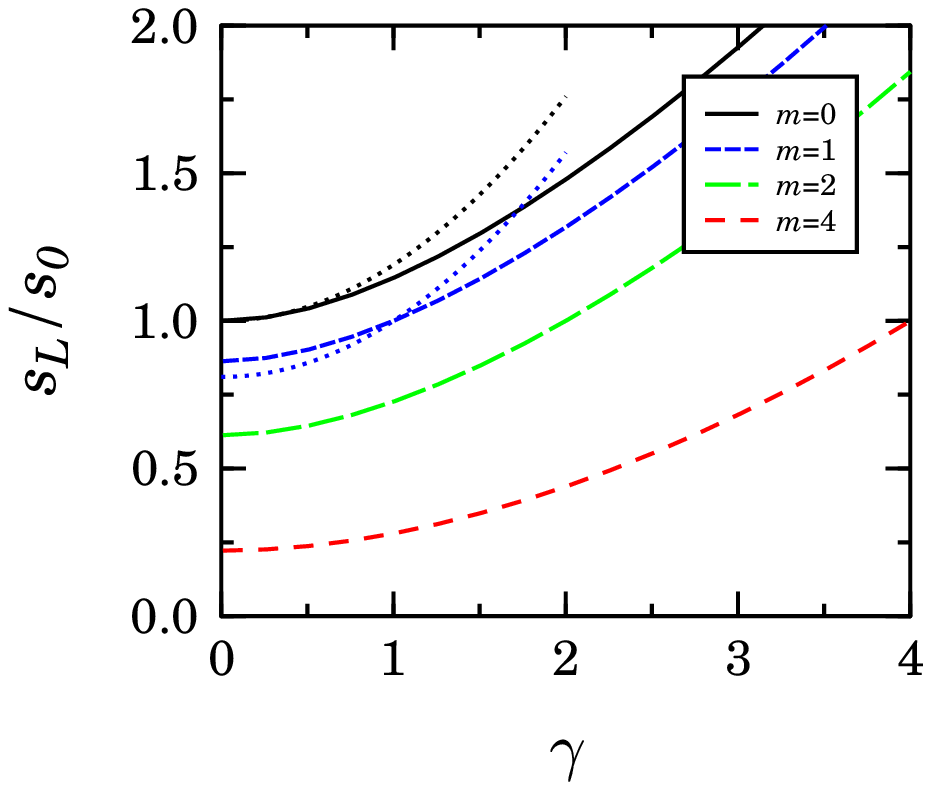}
 \hfill
 \raisebox{8mm}{
 \includegraphics[scale=0.55]{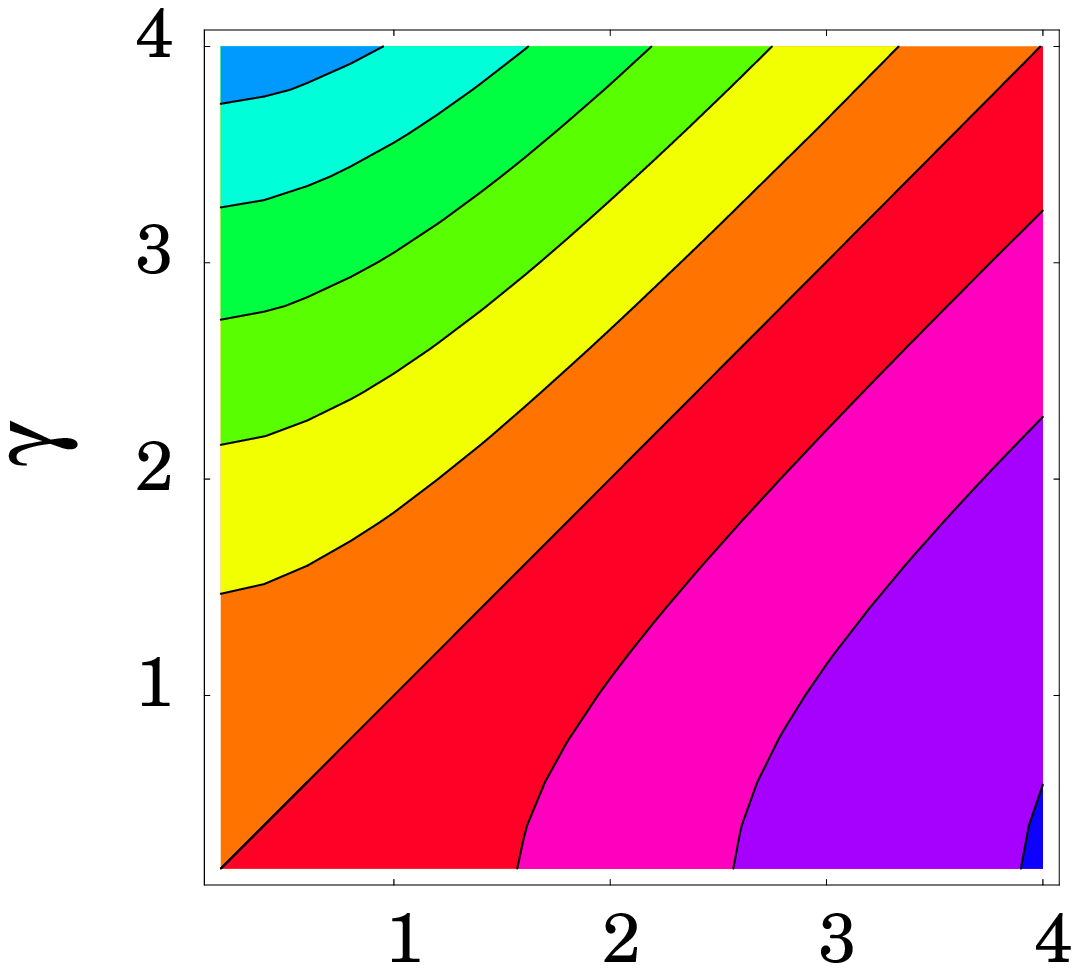}\rule{2em}{0em}}
 \caption{
   Left: The entropy $s_L(m,\gamma)$ as a function of $\gamma$ for 
   several values of $m$ ($\gamma$ and $m$ are in units of $T$).
   The dotted lines show the expansion (\ref{eq: sL expansion}) for $m=0$ 
   and $m/T=1$.
   Right: Contour plot of $s_L/s_0$; the contour spacing is
   0.25, and the straight line marks $s(m,\gamma)=s_0$.
  \label{fig: s(m,Gamma)}}
\end{figure}
An interesting detail is that $s_L(m=\gamma)$ is equal to the Stefan-Boltzmann entropy of the massless ideal gas, $s_0 = \frac4{90}\, \pi^2 T^3$. This is proven in the Appendix, where also the expansion 
\be
  s_L(m,\gamma)
  =
  s_0 \l[ 1 - \frac{15}{8\pi^2}\, \frac{m^2}{T^2}
          + \frac{15}{8\pi^2}\, \frac{\gamma^2}{T^2} + \ldots
      \r] 
 \label{eq: sL expansion}
\ee
for small values of $m$ and $\gamma$ is derived.
It is interesting to note that this result can be obtained by the expansion of the contribution $s_L^{(0)}$ with complex masses, $s_L(m,\gamma) \approx \frac12 (s_L^{(0)}(m+i\gamma)+s_L^{(0)}(m-i\gamma))$.

\subsection{Momentum-dependent mass and width parameters}
\begin{figure}[ht]
 \hskip -1em
 \includegraphics[scale=0.75]{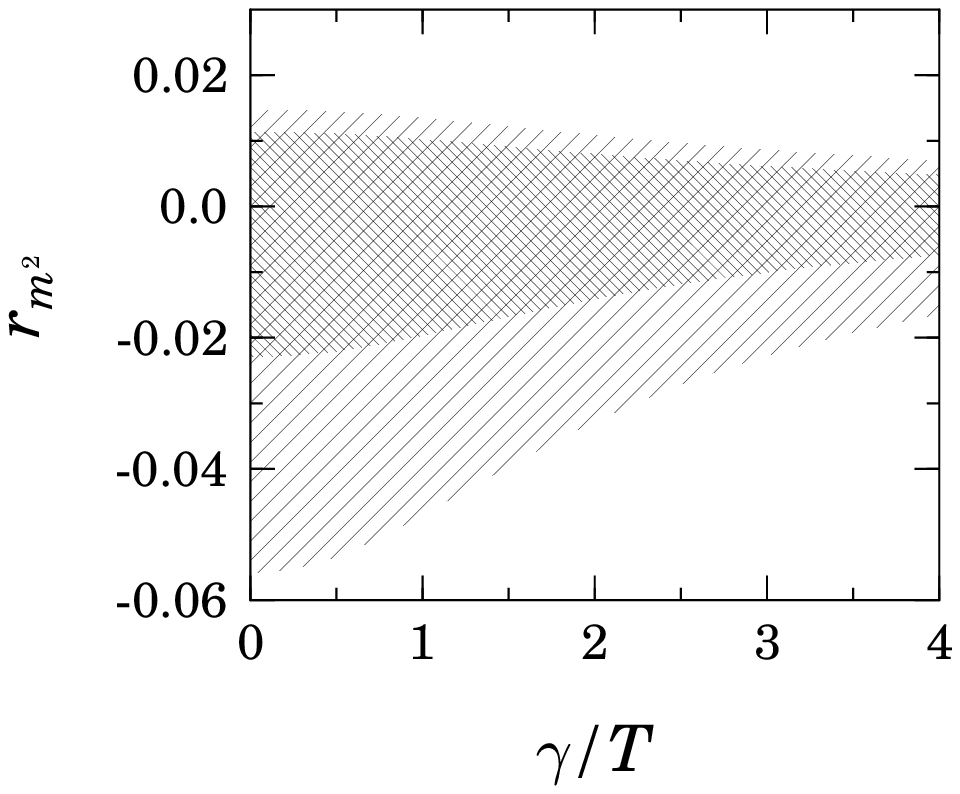}
 \hfill
 \includegraphics[scale=0.75]{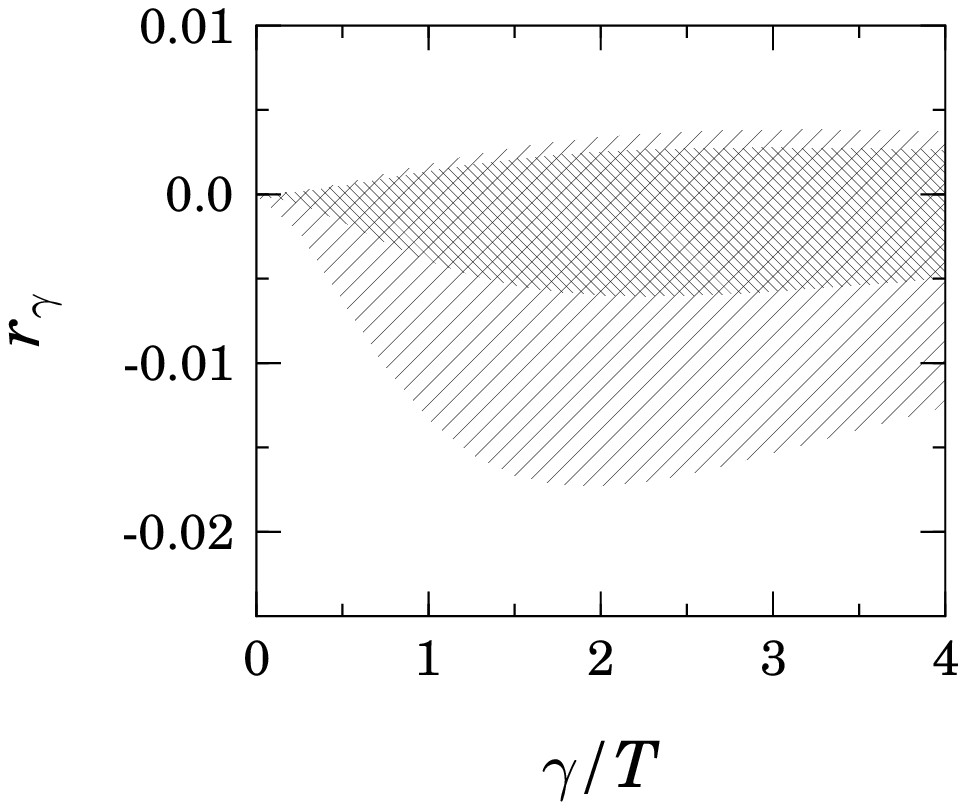}
 \caption{The sensitivity of the entropy $s_L$ on the momentum dependence 
    of $m^2$ (left) and $\gamma$ (right). 
    The parameters are varied for $k<T$ (see text) in the range 
    $[\frac14,4]$ (outer band) and $[\frac12,2]$ (inner band).
    In both cases the asymptotic mass is $m=4T$.
 \label{fig: r}}
\end{figure}

Due to phase space, thermodynamic bulk properties are mostly determined by hard momenta. Therefore, the entropy is expected to be not very sensitive on the exact momentum dependence of the width as well as on the dispersion relation (described by a momentum dependent mass parameter) at soft momenta.
To quantify this expectation, the squared mass and the width are varied for $k<T$ by some factor from $m^2$ and $\gamma$, which are now considered as the asymptotic values. Denoting the resulting entropy by $\tilde{s}_L$, the quantity
\[
  r = 1-\frac{\tilde{s}_L}{s_L} \, ,
\]
provides a measure of the momentum sensitivity.
As shown in Fig.~\ref{fig: r}, $r$ is indeed only of the order of a few percent when varying the dispersion relation.
The sensitivity to the low-momentum behavior of the width is even less.
The Figures show the sensitivity for a rather large asymptotic mass; for smaller masses the sensitivity is even lower. 
This quantifies the expectation that the entropy is, to the extent required below, insensitive to details of the propagator at soft momenta.

\subsection{Specific shape of spectral function}
For a first test of the sensitivity to the specific form of the spectral function, let's consider the normalized function
\be
  \rho_Q(\omega)
  =
  \frac{\sqrt2\gamma^3}E \l(
    \frac1{(\omega-E)^4+\gamma^4} - \frac1{(\omega+E)^4+\gamma^4}
  \r) ,
  \label{eq: rho_Q}
\ee
which has a more pronounced peak than the Lorentzian (\ref{eq: rhoL}).
It can be expressed in terms of the function $\rho_L$ with complex width parameters,
\be
  \rho_Q[\gamma]
  =
  \frac1{\sqrt2}
  \l(
    \sqrt{i}\,\rho_L[\sqrt{i}\gamma] + \sqrt{-i}\,\rho_L[\sqrt{-i}\gamma]
  \r) .
  \label{eq: rho_Qa}
\ee
An analogous relation easily allows to obtain the corresponding propagator in terms of the Lorentz propagator (\ref{eq: Delta_L}). Replacing furthermore $\gamma \rightarrow \sqrt2 \gamma$, the result reads
\be
  \Delta_Q(\omega,\bm k)
  =
  \frac{a^3+2ab^2-4ib^3}{a^4+4b^4} \, ,
  \label{eq: Delta_Q}
\ee
where $a = \omega^2-\omega_m^2$ and $b = 2\gamma \omega$. The propagator (\ref{eq: Delta_Q}) coincides with $\Delta_L(\omega,\bm k)$ for $\omega \rightarrow 0$ and $\omega \rightarrow \pm \infty$ as well as on the common mass shell $\omega^2 = \omega_m^2$. Thus, differences in the entropies can indeed be attributed to the spectral form rather than to a change in the dispersion relation.

\begin{figure}[hbt]
 \hskip -1em
 \includegraphics[scale=0.75]{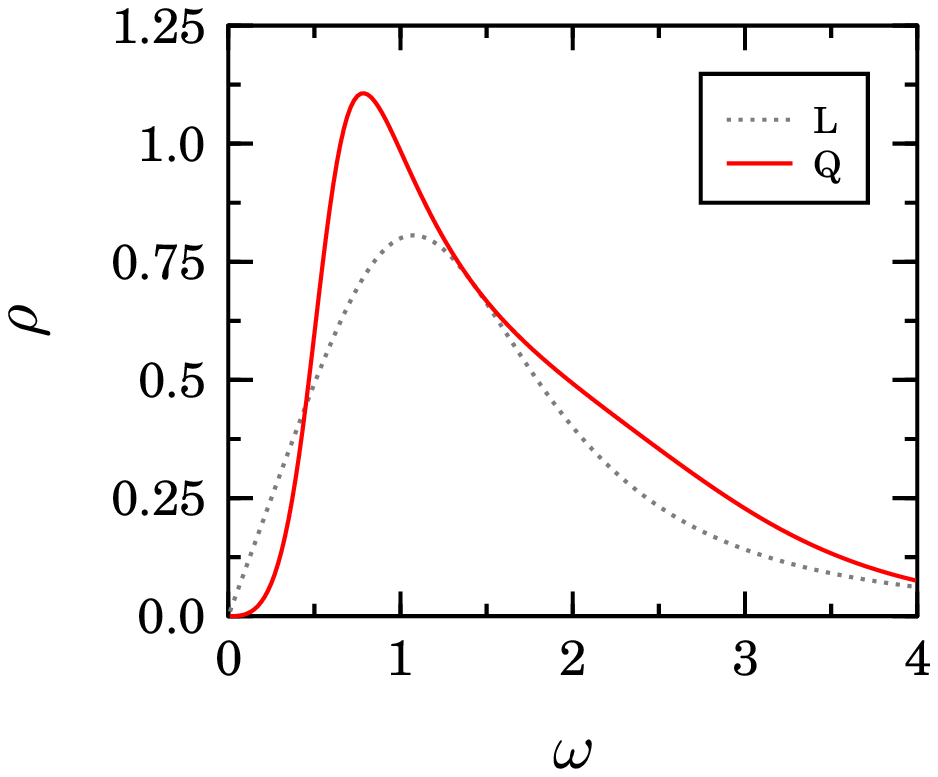}
 \hfill
 \includegraphics[scale=0.75]{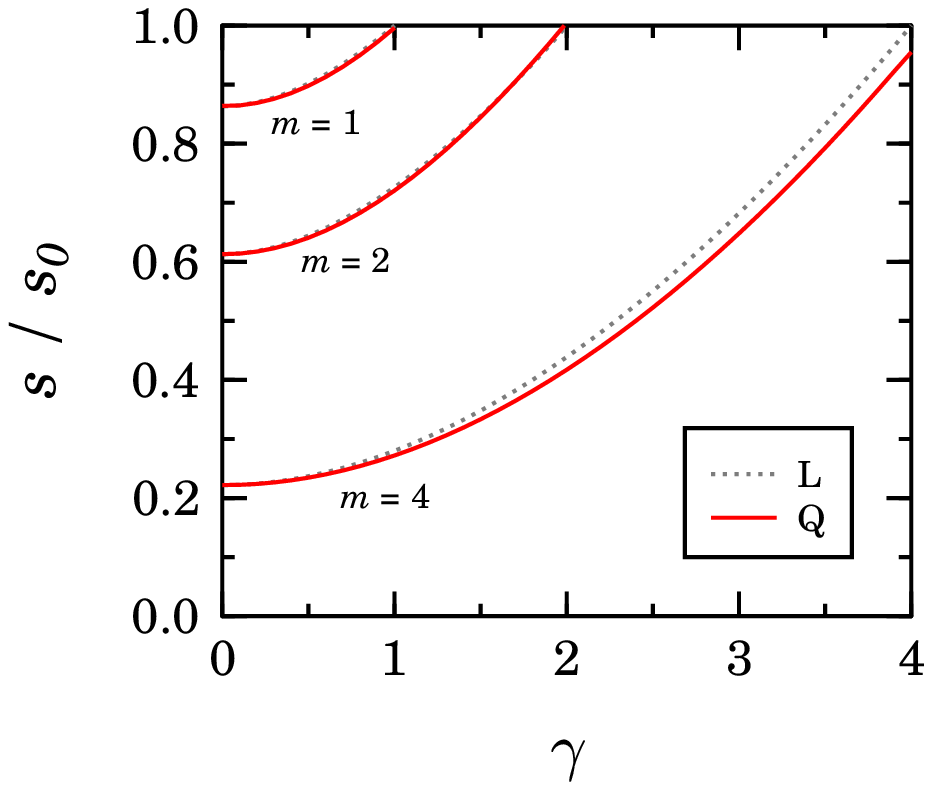}
 \caption{Left: The spectral functions of the propagators $\Delta_L$ and 
    $\Delta_Q$ with $\omega_m = \sqrt2$ (all quantities are in units of 
    $\gamma$); if $\gamma = T$, this corresponds, e.\,g., to $k=m=T$.
    Right: The corresponding entropies as functions of $\gamma$ and $m$ 
    (in units of $T$); for small masses the lines practically coincide.
 \label{fig: LvsQ}}
\end{figure}

The differences in the propagators $\Delta_L$ and $\Delta_Q$, apart from the analytic structure, are lucid for typical values of the 4-momentum, see Fig.~\ref{fig: LvsQ}. 
Nonetheless, there is almost no effect on the entropy even for large values of $m$ and $\gamma$. 
This example leads to the question which features of $\rho$ the entropy is actually sensitive to. Already expected in Sec.~\ref{sec 2} were differences between the two generic types of spectral functions shown in Fig.~\ref{fig: types of prop}. In any case, the integrand of the contribution (\ref{eq: Delta s}) is notably different from zero only for energies $|\omega-\omega_k| \ksim \gamma$, cf.\ Fig.~\ref{fig: intgrd_Delta_s}. A way to focus on this relevant interval is to consider large times, $t > \gamma^{-1}$, in the Fourier transform $\rho(t)$ of the spectral function.

\subsection{Spectral functions in Fourier space}
The Fourier transform of the spectral function is defined by
\be
  \rho(t)
  =
  \int_{-\infty}^\infty 
    \frac{d\omega}{2\pi}\, e^{-i\omega t}\, \rho(\omega) \, .
\ee
The sum rule (\ref{eq: sum rule}),
\[
  1
  =
  \int_{-\infty}^\infty \frac{d\omega}{2\pi}\, \omega\, \rho(\omega)
  =
  \int_{-\infty}^\infty dt\, \rho(t)   
  \int_{-\infty}^\infty \frac{d\omega}{2\pi}\,\omega\, e^{i\omega t} \, ,
\]
which after a partial integration becomes $i\int\!dt\,\dot\rho(t) \delta(t)$, then translates into
\be
  \l. \frac{d\rho(t)}{dt} \r|_{t=0} = -i \, .
  \label{eq: sum rule Fspace}
\ee
It is plausible that the sum rule tests the small-$t$ behavior in Fourier space since in momentum space it is closely related to the fact that the propagator approaches the free limit at large energies.
The Lorentzian spectral function (\ref{eq: rhoL}), with
\[
  \rho_L(t)
  =
  \exp(-\gamma|t|)\, \frac{\sin Et}{iE}\, ,
\]
obviously complies with the condition (\ref{eq: sum rule Fspace}).
From the fact that the oscillations and the attenuation of $\rho(t)$ are related to the position and the width, respectively, of the peak of $\rho(\omega)$, one can easily construct other conceivable spectral functions. In the ansatz
\be
  \rho_f(t)
  =
  f(t)\, \frac{\sin Et}{iE} \, ,
  \label{eq: rho models}
\ee
functions with $f(0)=1$ provide candidates for possible spectral functions if $\omega \rho(\omega) \ge 0$ is satisfied.

The similar entropies for the propagators $\Delta_L$ and $\Delta_Q$, cf.\ Fig~\ref{fig: LvsQ}, may now be attributed to the similar large-$t$ behavior of the spectral functions (basically also $\rho_Q(t)$ decreases exponentially, see Eq.~(\ref{eq: rho_Qa})).
Before studying this in more detail, it is noted that in terms of $\rho(t)$ the retarded propagator reads
\be
  \Delta(k_0)
  =
  i^{-1}\int_0^\infty dt\, e^{ik_0 t}\, \rho(t) \, .
  \label{eq: Delta from rho(t)}
\ee
For the ansatz (\ref{eq: rho models}),
\bea
  \Delta_f(k_0)
  &=&
  \frac{i}{2E} 
  \int_0^\infty dt\, e^{ik_0 t} \l( e^{iEt}-e^{-iEt}\r) f(t)
  \nonumber \\
  &=&
  \frac{i}{2E} \l( {\cal F}(k_0+E)-{\cal F}(k_0-E) \r) ,
\eea
the propagator can be expressed by a Fourier transform of the function $f(t)$,
\be
  {\cal F}(\omega)
  =
  \int_0^\infty dt\, e^{i\omega t} f(t) \, .
  \label{eq: Ftrafo}
\ee

\subsection{Non-exponential time behavior
\label{sec: sensitivity on rho}}
Although often assumed, an exponential decrease of $\rho(t)$ is not dictated by any fundamental requirement \cite{FettW}.
As already argued, the contribution (\ref{eq: Delta s}) to the entropy, due to the non-zero width, is determined by the long-time behavior of the spectral function; it will increase if $\rho(t)$ decreases faster, either by a larger value of $\gamma$ or due to the functional form of $\rho(t)$.
In the following, this is demonstrated systematically by some models for the spectral function (\ref{eq: rho models}), which are summarized in Table~\ref{tab: 1}.
\begin{table}[ht]
\centering{
\begin{tabular}{c|cc}
model & $f(t)$ & $\gamma {\cal F}(\omega)$\\
\hline
\\[-0.7em]
$L$    & $\exp\l( -\gamma|t| \r)$  & $(1-ix)^{-1}$
\\[0.5em]
$P_1$  & $(1+\gamma |t|)^{-1}$ & $e^{-ix}\Gamma(0,-ix)$ 
\\[0.5em]
$P_2$  & $(1+\gamma|t|)^{-2}$ & $1+ixe^{-ix}\Gamma(0,-ix)$ 
\\
\vdots & \vdots & \vdots
\\
$P_n$  & $(1+\gamma|t|)^{-n}$ & $(1+ix[\gamma{\cal F}_{n-1}])/(n-1)$
\\[0.7em]
$P_1^\star$  & \quad$(1+(\gamma t)^2)^{-1/2}$ & 
        \quad$K_0(|x|)+\frac\pi2 {\rm sgn}(x)[I_0(|x|)-L_0(|x|)]$
\\[0.7em]
$G$ &  $\exp\l(-(\gamma t)^2\r)$ & 
        \quad$\frac12\sqrt\pi \exp\l(-(x/2)^2\r) \l[ 1+{\rm Erf} (ix/2)\r]$
\end{tabular}}
\caption{Damping models (\ref{eq: rho models}) for $\rho(t)$, together 
    with their Fourier transforms ${\cal F}$ defined in 
    Eq.~(\ref{eq: Ftrafo}).
    $\Gamma(0,z)$ denotes the incomplete $\Gamma$-function, $K_0$ is the 
    modified Bessel function of the second kind, $L_0$ is the modified 
    Struve function, Erf is the error function, and $x = \omega/\gamma$.
  \label{tab: 1}}
\end{table}

\begin{figure}[ht]
\begin{center}
 \hskip -1em
 \includegraphics[scale=0.75]{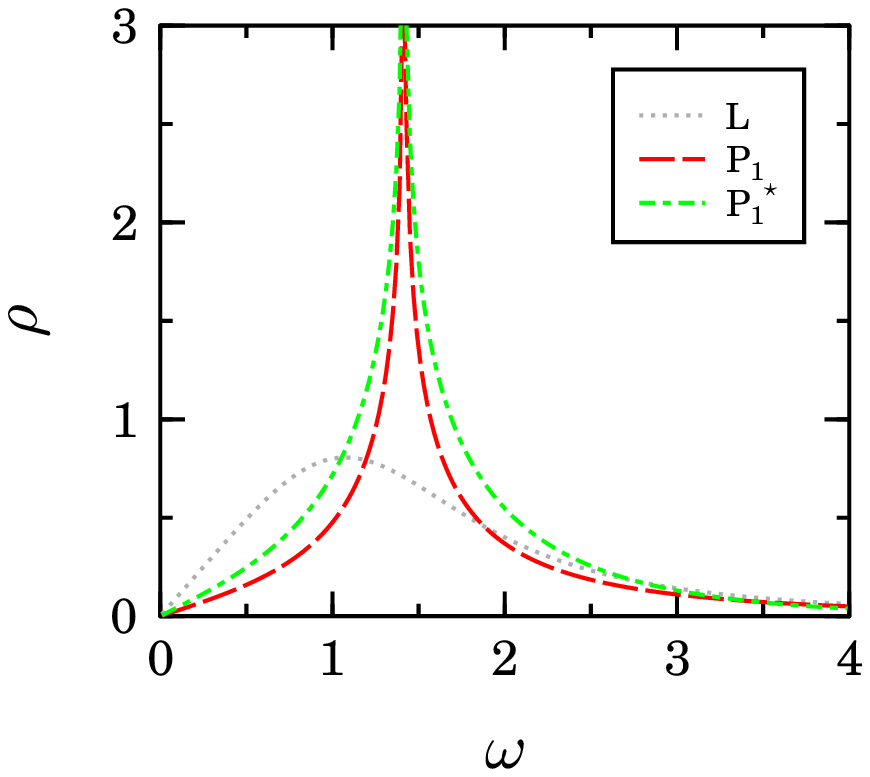}
 \hfill
 \includegraphics[scale=0.75]{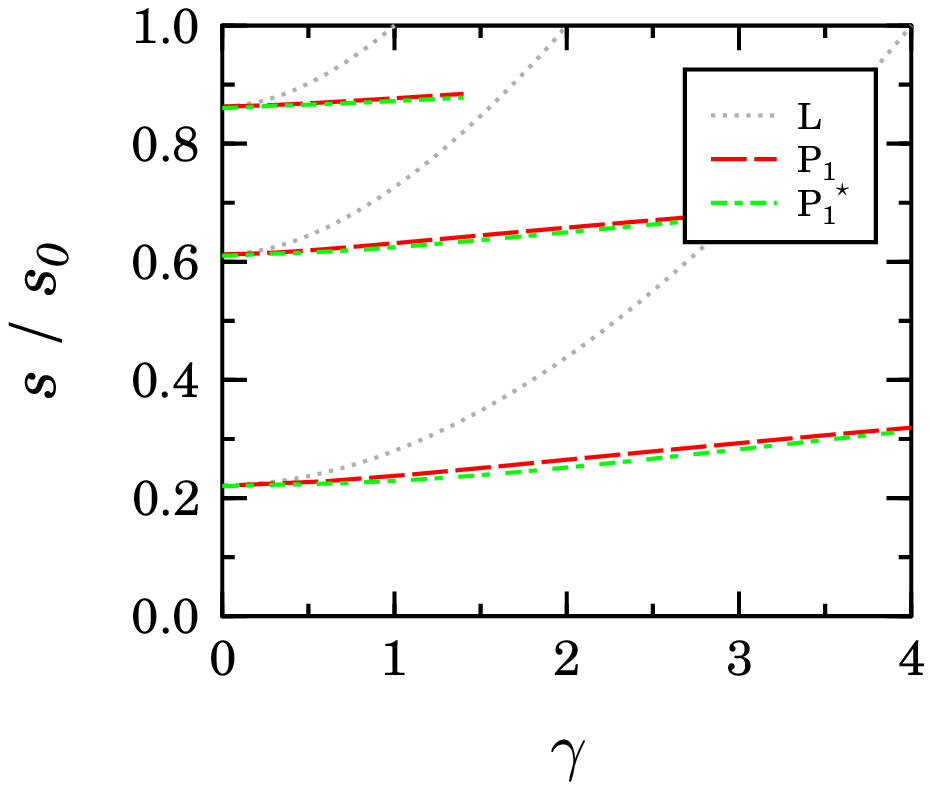}
\end{center}
 \caption{Comparison of the models $P_1$ and $P_1^\star$ with $L$ 
    analogous to Fig.~\ref{fig: LvsQ}. The entropies almost coincide
    for the cases $P_1$ and $P_1^\star$.
 \label{fig: Lvs1}}
\end{figure}

In the models $P_1$ and $P_1^\star$, $\rho(t) \sim f(t)$ decreases asymptotically as $t^{-1}$. This implies that the spectral function
\be
  \rho(\omega)
  =
  -2{\rm Im}\Delta(\omega)
  =
  -\frac1E\, \l( {\cal F}_c(\omega+E)-{\cal F}_c(\omega-E) \r) ,
\ee  
where ${\cal F}_c(\omega) = \int_0^\infty dt \cos\omega t\, f(t)$ is the cosine transform of $f(t)$, diverges logarithmically at $\omega = \pm E$, see Figs.~\ref{fig: types of prop} and \ref{fig: Lvs1}.
For $\gamma \rightarrow 0$, the free propagator is recovered. 
For the model $P_1$, e.\,g., this follows from $\Gamma(0,x) = e^x(x^{-1} + \ldots)$ for $x \rightarrow \infty$.
Moreover, the incomplete $\Gamma$-function $\Gamma(0,z)$ is discontinuous at the negative axis. Accordingly, the retarded propagator $\Delta_{P1}$ has cuts in the lower $k_0$-plane, starting at the singularities.
The entropies for the models $P_1$ and $P_1^\star$ are shown in Fig.~\ref{fig: Lvs1}.
As anticipated before, the entropy increases with $\gamma$ much less than in the case of a regular spectral function.
The deviations between the models $P_1$ and $P_1^\star$ are only at the level of a few percent. This demonstrates that the entropy is indeed sensitive to the large-$t$ behavior of $\rho(t)$, while the short-time behavior of the models is irrelevant.

\begin{figure}[ht]
\begin{center}
 \hskip -1em
 \includegraphics[scale=0.75]{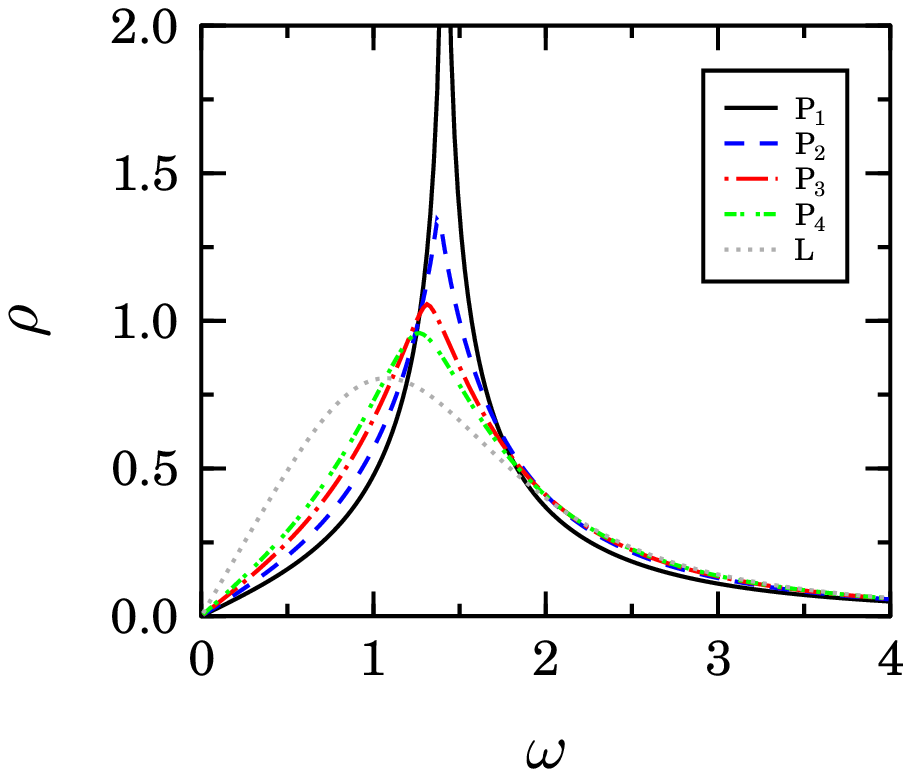}
 \hfill
 \includegraphics[scale=0.75]{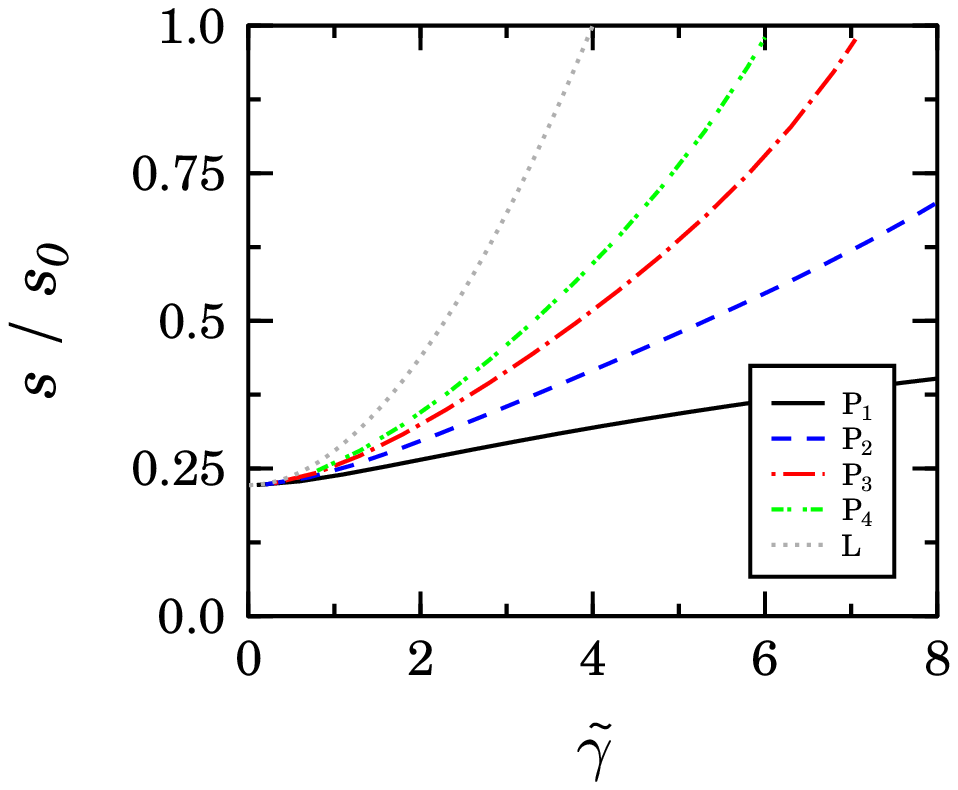}
\end{center}
 \caption{Results for the models $P_n$ ($n=1...4$) similar to 
    Fig.~\ref{fig: LvsQ}. The left plot compares the spectral functions 
    $\rho_n$ with a width parameter $\gamma_n = \tilde\gamma/n$ and 
    $\rho_L$ with $\gamma = \tilde\gamma = 1$; the right plot shows the 
    entropies as a function of $\tilde\gamma$.    
 \label{fig: Pn}}
\end{figure}

\begin{figure}[ht]
\begin{center}
 \hskip -1em
 \includegraphics[scale=0.75]{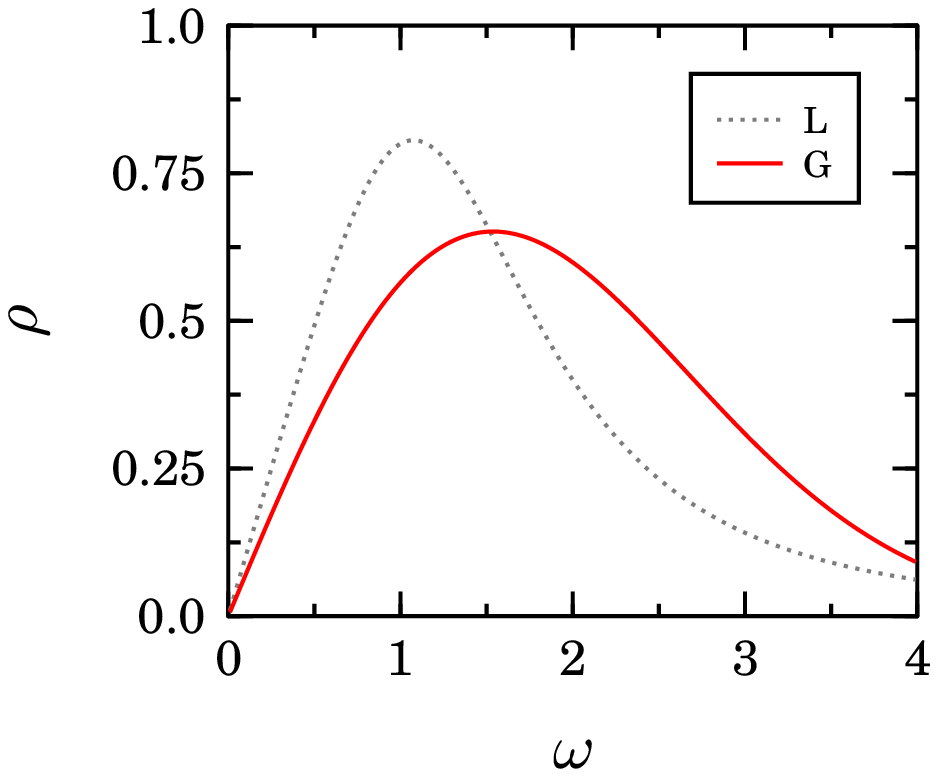}
 \hfill
 \includegraphics[scale=0.75]{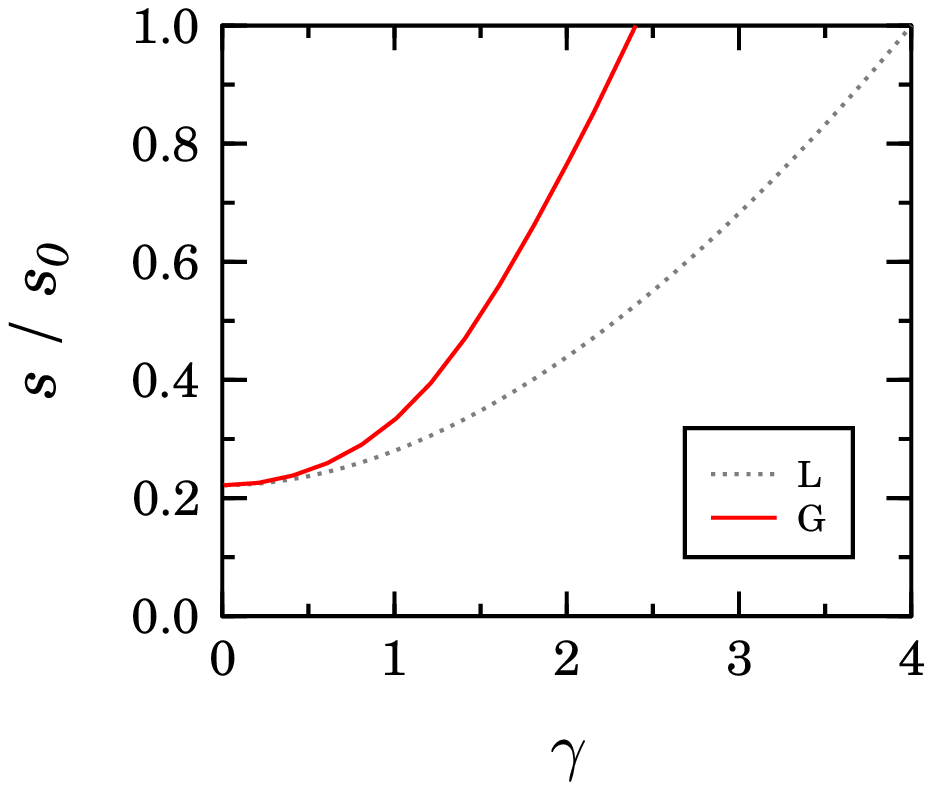}
\end{center}
 \caption{Results for the Gaussian model $G$ analogous to 
    Fig.~\ref{fig: LvsQ}.
 \label{fig: LvsG}}
\end{figure}

In the models $P_n$, the spectral function decreases as $t^{-n}$.
Since the functions $f_{Pn}(t)$ are related by derivatives with respect to $t$, their Fourier transform (\ref{eq: Ftrafo}) can be calculated by a simple recursion relation given in Table~\ref{tab: 1}.
Fig.~\ref{fig: Pn} shows the spectral functions for $n=1 \ldots 4$
along with the corresponding entropies.
Plotting $s$ as a function of $\tilde\gamma = n \gamma$ confirms again the general expectation that the entropy is not determined by the small-$t$ behavior, which is here $\rho_n(t) \sim 1-\tilde\gamma t$, but rather by the large-time behavior of the spectral function.
As expected, the faster $\rho(t)$ decreases, the larger the entropy.
Leaving the class of polynomial models, this trend is also obvious when considering the Gaussian model $G$ in Fig.~\ref{fig: LvsG}.

\section{QCD \label{sec 4}}
In QCD, the entropy for various numbers of quark flavors, plotted as a function of $T/T_c$ and scaled by the free limit, has a remarkably universal behavior as found in lattice calculations \cite{KarscLP}.
I focus here on the representative case of the quenched limit of QCD, and point out briefly expected differences for the physical case.

The gluon propagator has a transverse and a longitudinal part which leads to a corresponding decomposition of the entropy. The contributions have the form (\ref{eq: s dqp}) multiplied by the respective degeneracies \cite{BlaizIR}.
The longitudinal modes are collective excitations whose spectral strength is exponentially suppressed for larger momenta, which leads to only a minor contribution to the entropy. In the perturbative limit, it is of the order $g^3$ while the transverse modes yield a ${\cal O}(g^2)$ term. Also for larger coupling the longitudinal contribution is rather small as demonstrated in the HTL calculations \cite{BlaizIR, Peshi}, which actually might overestimate the effect since it leads to a negative total entropy \footnote{
    This is due to the negative definite ghost contribution which in 
    Ref.~\cite{BlaizIR} is implicitly included in the longitudinal 
    contribution.} 
at very large $g$. 
Taking therefore into account only the dominating transverse excitations with the propagator $\Delta$, the resummed entropy reads
\be
  s 
  = 
  -2(N_c^2-1) 
  \int_{k^4} \frac{\partial n}{\partial T}\,
   \l( \Im\ln(-\Delta^{-1}) + \Im\Pi\, \Re\Delta \rule{0em}{1.2em} \r) .
 \label{eq: s_SCQ}  
\ee
In a self-consistent approximation, the propagator will have a residual gauge dependence, leading to an unphysical result for the entropy. However, as motivated before, a parameterization of the exact propagator by the dispersion relation and the width, which are gauge invariant, can be used in (\ref{eq: s_SCQ}).

\begin{figure}[ht]
 \centerline{\includegraphics[scale=0.9]{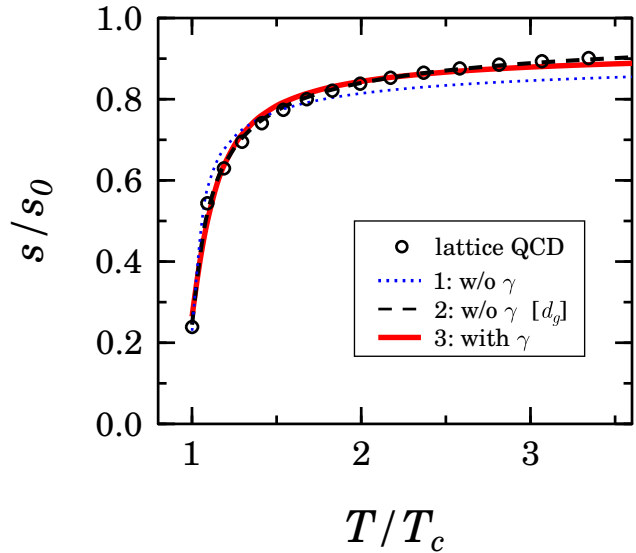}}
 \caption{The entropy of the SU(3) plasma in units of the free entropy.
    The symbols represent the lattice data \cite{CPPACS}.
    Fits without width are depicted by the dotted and dashed line, for
    the latter $d_g$ was also fitted (see text).
    The full line is the Fit 3 with the width.
    The parameters are summarized in Table~\ref{table}.
  \label{fig: s_fit}}
\end{figure}

The ansatz of the quasiparticle models \cite{pQP} is to neglect the width altogether, and to describe the propagator simply by the perturbative asymptotic self-energy on the light-cone, which is a gauge invariant mass term,
\be
  M^2 = \frac{N_c}6\, g^2 T^2 \, ,
 \label{eq: M2}
\ee
where $N_c=3$ \footnote{
    Also in the HTL calculations \cite{BlaizIR,Peshi}, this asymptotic mass
    is the only scale in the propagators, which have zero width as well but
    include a Landau-damping part.}.
This `minimal' resummation of the entropy can indeed nicely describe the lattice data for all temperatures above $T_c$ if an IR enhancement of the running coupling is permitted, for example in the form
\be
  g^2(T) = \frac{48\pi^2}{11N_c \ln(\lambda(T-T_s)/T_c)^2} \, .
 \label{eq: g}
\ee
For the physical number of degrees of freedom, $d_g = 2(N_c^2-1)$, a fit of the parameters $\lambda$ and $T_s/T_c$ leads in the entropy to small but systematic deviations from the lattice result for $T > 2T_c$, cf.\ Fit 1 in Fig.~\ref{fig: s_fit}. 
This can be improved when considering $d_g$ as an additional fit parameter, which yields a value not too far from the physical one. 
\begin{table}[ht]\centerline{
 \begin{tabular}{c || c | c | c | c}
      &$\quad\lambda\quad$&$\;T_s/T_c\;$&$\qquad d_g\qquad$&$\quad c\quad$
 \\ \hline
 Fit 1  &   10.5      &   0.88      &  16 (fixed) &    -
 \\ \hline
 Fit 2  &      5.2    &   0.76      &    17.5     &    -
 \\ \hline
 Fit 3  &    2.6      &   0.50      &  16 (fixed) &   14.0
 \end{tabular}}
 \caption{The parameters of the fits shown in Fig.~\ref{fig: s_fit}.
  \label{table}}
\end{table}
\begin{figure}[ht]
 \centerline{\includegraphics[scale=0.9]{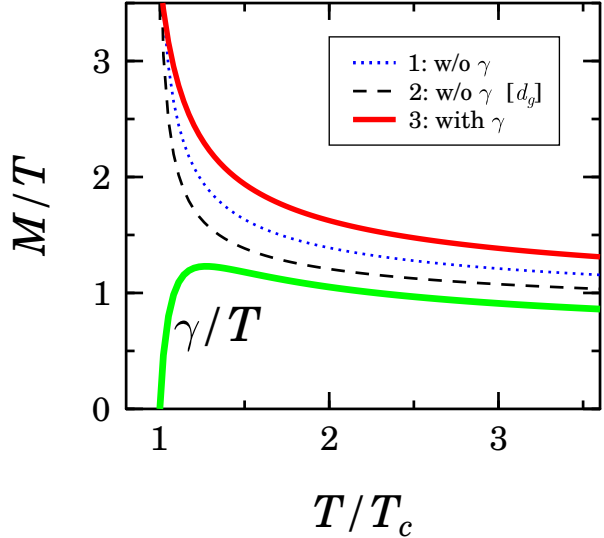}}
 \caption{The masses and the width according to the fits shown in 
    Fig~\ref{fig: s_fit}.
  \label{fig: M_gamma_fit}}
\end{figure}
While in the first reference \cite{pQP} the longitudinal modes were considered as one possible explanation, it does not seem likely after the longitudinal contribution was found to be negative in the HTL calculation \cite{BlaizIR}.

\begin{figure}[ht]
 \centerline{\includegraphics[scale=0.9]{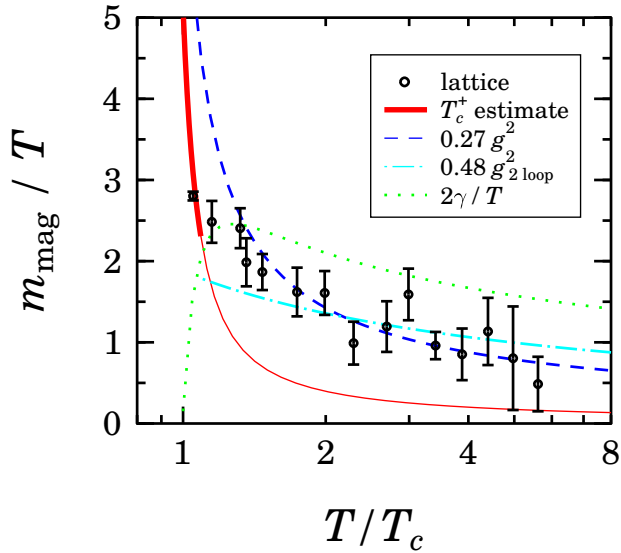}}
 \caption{The lattice data \cite{NakaSS} for the magnetic gluon mass, and
    the estimate based on (\ref{eq: gamma}) with the fitted width, which 
    is meaningful only for $T = T_c^+$, i.\,e., only for the left data 
    point (see text). 
    Also shown are the $g^2 T$-fits with the 2-loop running coupling 
    \cite{NakaSS} (dash-dotted line), and with the adjusted coupling 
    (\ref{eq: g}) from Fit 3 (dashed line). 
  \label{fig: m_mag}}
\end{figure}

Taking now into account the width on the same footing as the mass (\ref{eq: M2}), I will first consider a perturbative ansatz for $\gamma$. In the weak coupling limit, the width of a hard transverse gluon was obtained by Pisarski \cite{Pisar93} in a resummed calculation, as
\be
  \gamma
  =
  \frac{N_c}{8\pi}\, g^2T \l( 
   \ln\frac{\frac23\, M^2}{m_{\rm mag}^2+2m_{\rm mag}\gamma} + 1.09681...
  \r).
 \label{eq: gamma}
\ee
Several assumptions have been made here: ($i$) soft gluons are HTL dressed while intermediate hard gluons have a Lorentzian spectral function; ($ii$) the divergence from the static magnetic gluons is screened assuming that this sector of QCD can be  parameterized by a mass $m_{\rm mag} \sim g^2 T$. 
The first supposition follows the concept of a self-consistent calculation, hence $\gamma$ appears also on the right hand side of Eq.~(\ref{eq: gamma}) as a regulator next to the magnetic mass. Since the argument of the logarithm is basically $g^{-2}$, the width of moving excitations is enhanced compared to the width at rest, $\gamma(0) \sim g^2 T$.
Implicit with ($i$) is the supposition of a simple pole structure of the propagator, which is not warranted by any fundamental requirement. In fact, the result (\ref{eq: gamma}) could only be justified, to separate the pole from a branching point, if $\gamma \ll m_{\rm mag}$ (although physically the converse relation was considered more likely).
While the constraint was necessary to explore details of the cut-off, the generic behavior $\gamma \sim g^2 \ln(1/g^2) T$ is expected on general grounds \cite{Pisar89}.
To keep the connection to the result (\ref{eq: gamma}), I analyze the SU(3) entropy with the Lorentz spectral function and the width in the form \footnote{
    I mention that although Pisarski considered massless hard gluons, his 
    result also holds true for small masses $\sim gT$ due to a 
    cancellation in the energy difference of the inner and outer gluon.
    Note also that the resummation of a width $\sim g^2 \ln(1/g) T$ can
    generate powers of logs in the expansion of thermodynamic quantities, 
    cf.\ Eq.~(\ref{eq: sL expansion}).
    \label{footnote}}
\be
  \gamma
  =
  \frac3{4\pi}\, \frac{M^2}{T^2} \, T \ln\frac{c}{(M/T)^2} \, ,
 \label{eq: gamma fit}
\ee
where $c$ parameterizes the soft cut-off. It is emphasized that because the functional relation between $M$ and $\gamma$ is fixed, it is not obvious 
whether a fit is possible at all, or that introducing $c$ in addition to the parameters in the coupling (\ref{eq: g}) will improve the result.
The numerics shows, however, that this is indeed the case. 
The enhanced entropy, due to the width, nicely explains the small deviations of Fit 1 for $T \ge 2$ without having to adjust $d_g$. As to be expected, the mass is somewhat larger than in the previous fits, cf.\ Fig.~\ref{fig: M_gamma_fit}. More interesting is the behavior of the width near $T_c$. Because there $s/s_0$ is small, the mass and hence the coupling have to be large. At the same time, the width cannot be too large since it would over-compensate the decreasing effect of the mass. Within the ansatz (\ref{eq: gamma fit}) this implies that the logarithm has to become small. It is worth to emphasize that the optimal value of $c$, given in Table~\ref{table}, is surprisingly close to the value
\be
  c^\star = \frac{M^2(T_c)}{T_c^2} \approx 13.7 \, ,
\ee
so the width vanishes almost precisely at $T_c$ (and is indeed positive for all temperatures) \footnote{
    Since the value of $c$ may appear to be closely related to the small 
    entropy at $T_c$ it is noted that it is in fact sensitive to the 
    global behavior of $s(T)$.}.
Interestingly then, the condition $\gamma \ll m_{\rm mag}$, which was necessary in the derivation of Eq.~(\ref{eq: gamma}) (but originally considered to not represent the physical situation), can actually be fulfilled in a small vicinity, $T_c^+$, of $T_c$. Taking the next-to-leading logarithm result at face value leads to an estimate of the magnetic mass which is consistent below $1.1T_c$, where $m_{\rm mag} > 2\gamma$.
This estimate (including the `predicted' range of applicability) is indeed in nice agreement with the lattice data \cite{NakaSS} as shown in Fig.~\ref{fig: m_mag}. 
The magnetic mass at $T_c$, which is difficult to calculate on the lattice, is estimated as
\be
  m_{\rm mag}(T_c) 
  \approx 
   \sqrt{\frac2c}\, \frac{M(T_c)^2}{T_c} 
  \approx 5 T_c \, .
  \label{eq: m_mag}
\ee
For larger temperatures, the behavior of the magnetic mass is consistent with the expectation $m_{\rm mag} = d\, g^2 T$. Note that the fit from Ref.~\cite{NakaSS}, which uses the 2-loop running coupling at the momentum scale $2\pi T$, can be improved by using the coupling as obtained in Fit~3.

It is emphasized that the magnetic and the electric screening masses have a qualitatively different large-coupling behavior: The electric screening mass, as known from lattice calculations \cite{NakaSS}, becomes {\em small} near $T_c$. This is readily understood in the quasiparticle model \cite{pQP}, where
\be
  m_{\rm D}^2 
  =
  \Pi_{00}(\omega=0, \bm p\rightarrow 0) 
  =
  -2g^2 N_c \int_{k^3} \l.\frac{\p n}{\p\omega}\r|_{\omega_m} \, .
\ee
Near $T_c$, the excitations are narrow (thus the quasiparticle picture is justified) and heavy, and the Debye mass
\be
  m_{\rm D}^2 
 \sim
 g^2\, g^{3/2} e^{-M/T}\, T^2
\ee
is Boltzmann-suppressed.
This decrease cannot be expected, even tendentiously, from the next-to-leading order perturbative result \cite{Rebha},
\be
  m_{\rm D, n}^2 
  =
  m_{\rm D, 0}^2 \l[
    1+\frac{\sqrt{3N_c}}{2\pi}\, g 
        \l( \ln\frac{2m_{\rm D, n}}{m_{\rm mag}} -\frac12 \r)
  \r] .
\ee
Although apart from the more obvious solution, which is larger than $m_{\rm D, 0} = (N_c/3)^{1/2} gT$, there is a second smaller solution (at $T_c$, with (\ref{eq: m_mag}), it is $\bar{m}_{\rm D, n} \approx 0.4 m_{\rm D, 0}$), the latter is unphysical because it does not approach the correct perturbative limit.

Coming back to the discussion of the width, it is plausible from the properties of the entropy (e.\,g., from the fact $s_L(m=\gamma)=s_0$ mentioned in Sec.~3) that the width has to be rather small near $T_c$.
However, the functional form (\ref{eq: gamma fit}), even as an extrapolation of the perturbative form similar to (\ref{eq: M2}), is a priori not justified near $T_c$ where $M$ becomes large (see footnote~[37]). %\ref{footnote}). 
Physically, one would here expect again a Boltzmann suppression of the heavy thermal fluctuations, hence
\be
  \bar\gamma = A e^{-bg}\, g^\nu\, T \, .
 \label{eq: gamma fit2}
\ee
For lack of better knowledge, I take this generic form as an ansatz for large coupling, and smoothly connect it to the `perturbative' form (\ref{eq: gamma fit}) with the adjusted values of $\{\lambda, T_s/T_c, c\}$, by
\[
  \gamma^\star
  =
  (1-\Theta)\bar\gamma + \Theta\gamma \, ,
\]
with $\Theta(T) = \frac12 + \pi^{-1} \arctan((T-\bar{T})/\delta)$.
Since the fit function $s/s_0$ can basically be described by 3 parameters (say by the values at $T_c$ and in the saturation-like regime, and by the slope at $T_c$), a conclusive determination of the parameters $\{A,b,\nu,\bar{T},\delta\}$ cannot be expected. In any case, within the enlarged parameter space the improvements in $\chi^2$ compared to Fit 3 are only of the order of a few percent; the changes in the plot of the mass and the width, including the distinguished behavior at $T_c$, are almost invisible. This robustness of the results justifies a posteriori the usage of the perturbative ansatz (\ref{eq: gamma fit}) also for smaller temperatures.

\begin{figure}[ht]
 \hskip -1em
 \centerline{
    \includegraphics[angle=-90,scale=0.5]{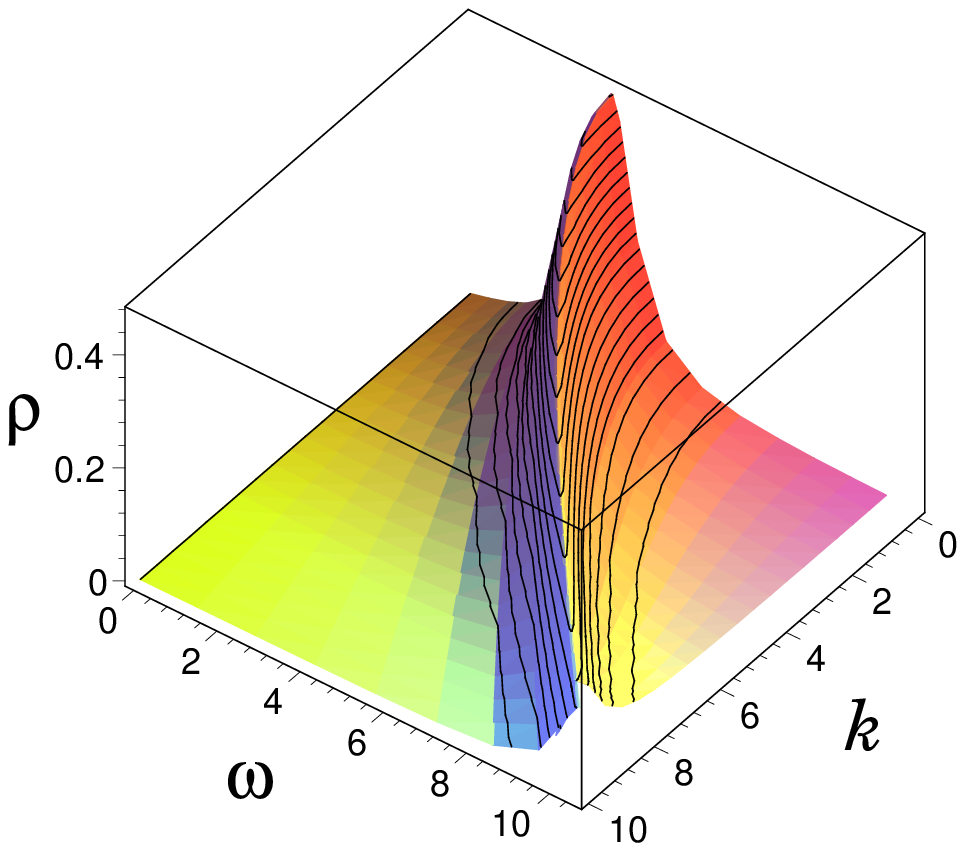}\hskip 2em
    \includegraphics[angle=-90,scale=0.5]{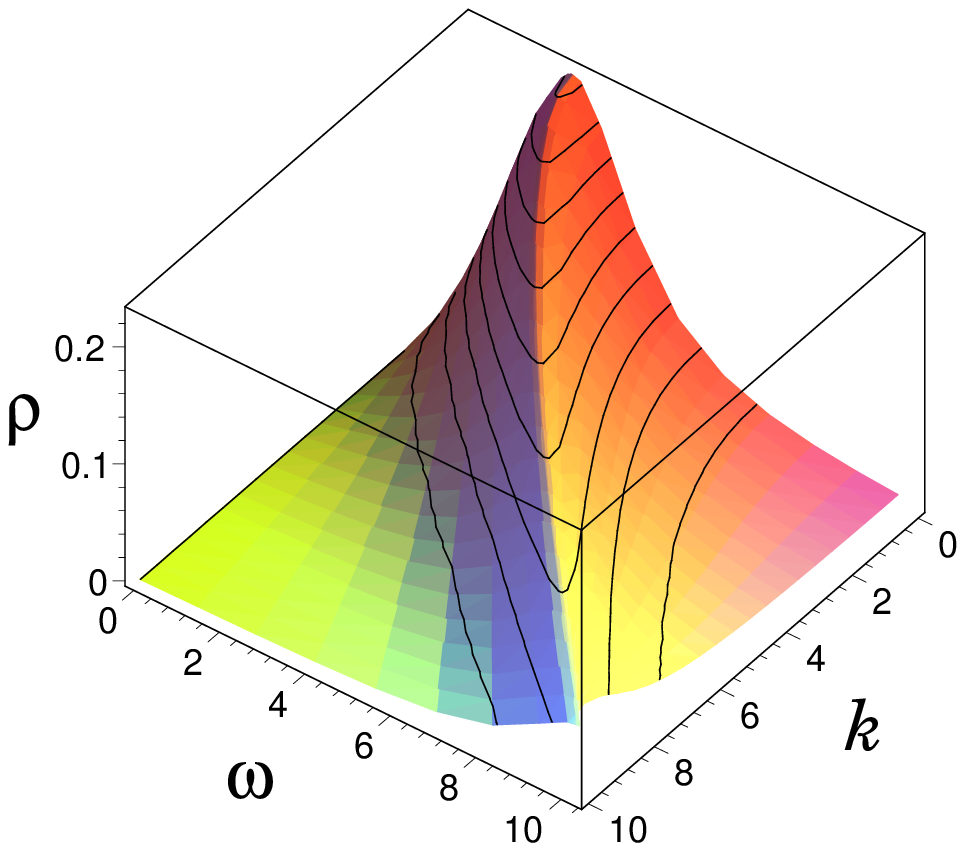}\hskip 2em
    \includegraphics[angle=-90,scale=0.5]{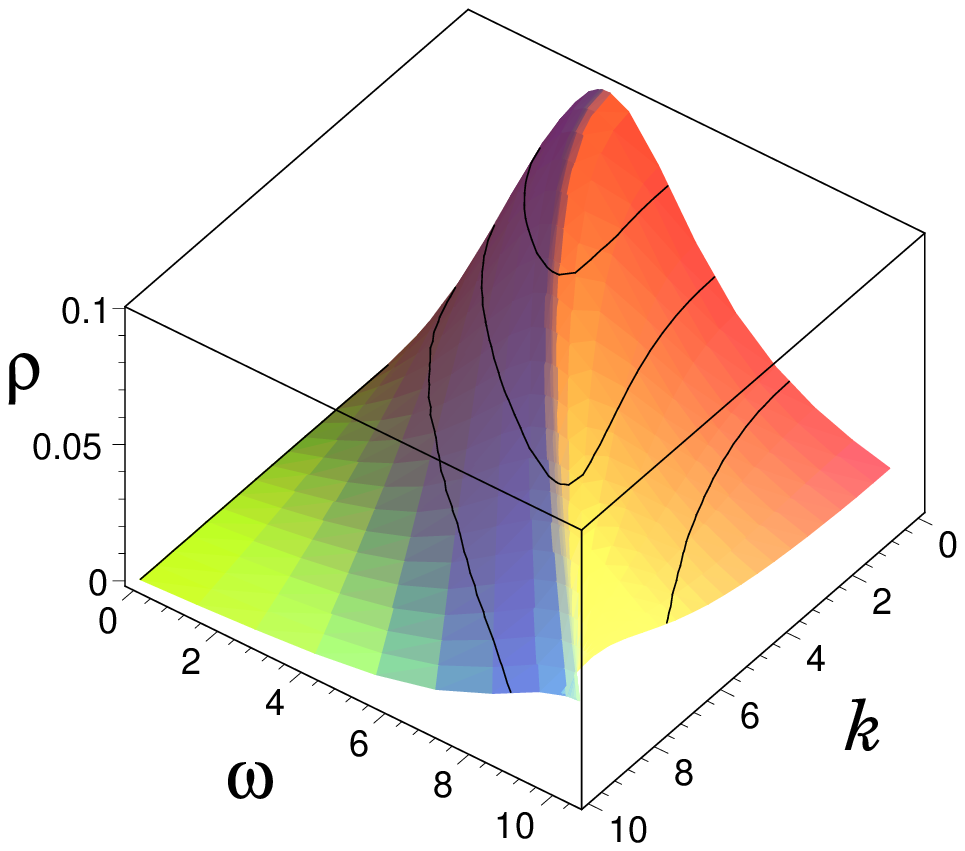}}
 \vskip 1em  
 \caption{The Lorentzian gluon spectral function from Fit 3 for 
    $T/T_c = 1.03,\ 1.35,\ 3$ ($\omega$, $k$ and $\rho$ are in units 
    of $T$).
    Shown here is the full phase space although the present approach can 
    make statements only for hard momenta, of the order of $T$ or larger.
  \label{fig: rho_fit}}
\end{figure}

For larger $T$, after a distinguished maximum at 
\be
  T_\gamma \approx 1.3 T_c \, ,
\ee
the ratio $\gamma/T$ decreases very slowly. Different from what the parametric form of Eq.~(\ref{eq: gamma fit}) might suggest, the width is even for rather large $T$ to a good accuracy proportional to the mass; for $T/T_c$ in $[5,100]$, 
\be
  \frac\gamma M \approx 0.69 - 0.02 \ln\frac{T}{T_c} \, .
\ee
This underlines the fact that in this range of temperatures quasiparticle models can provide only an effective description, while making a relation to the actual excitations may be difficult. Near $T_c$, on the other hand, the transverse hard excitations can be directly interpreted as quasiparticles as apparent from Fig.~\ref{fig: rho_fit}.
This concept is beneficial since (up to rather large temperatures) the system is still strongly coupled: terms of higher order in $g$ contribute significantly in the resummed entropy.

For definiteness, I have considered for QCD the case of a Lorentz spectral function. However, from the results of Sec.~\ref{sec: sensitivity on rho} it is obvious that the main result -- a small width near $T_c$ -- should hold true also for other spectral functions, unless their Fourier transform has an exotic long-time behavior such as $\rho(t) \sim t^{-1}$, which is not to be expected.

\section{Conclusions}

In this work it was shown for the deconfined SU(3) plasma that the collisional width (or damping rate) of hard gluons should be sizeable at intermediate temperatures, but has to become small near $T_c$. While from an extrapolation of the parametric estimate $\gamma \sim g^2\ln(1/g)T$ (with the logarithm read as an enhancement factor) this result may seem surprising, a large width would be hard to reconcile with the small entropy near $T_c$ as established in lattice calculations. 
For QCD with dynamical quarks, the rescaled entropy has a similar temperature dependence \cite{KarscLP}. Therefore, the principal result of a small width near the transition carries over from the quenched to the physical case. Although in full QCD $s/s_0$ is slightly larger at $T_c$, which is related to the nature of the transition, and the widths of the hard transverse gluons and quark particle-excitations might not fully vanish, a quasiparticle picture of the strongly coupled QCD plasma close to $T_c$ appears justified. 

\begin{figure}[ht]
 \centerline{\includegraphics[scale=0.9]{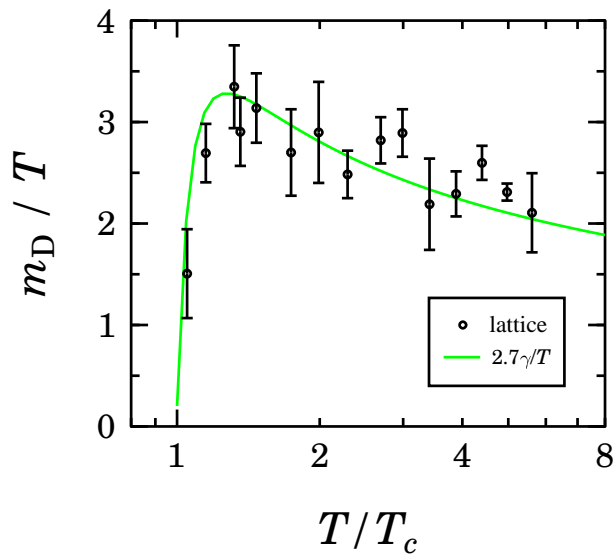}}
 \caption{The lattice data \cite{NakaSS} for the gluon Debye mass and the 
    rescaled width from Fit 3.
  \label{fig: m_debye}}
\end{figure}

\begin{figure}[ht]
 \centerline{\includegraphics[scale=0.9]{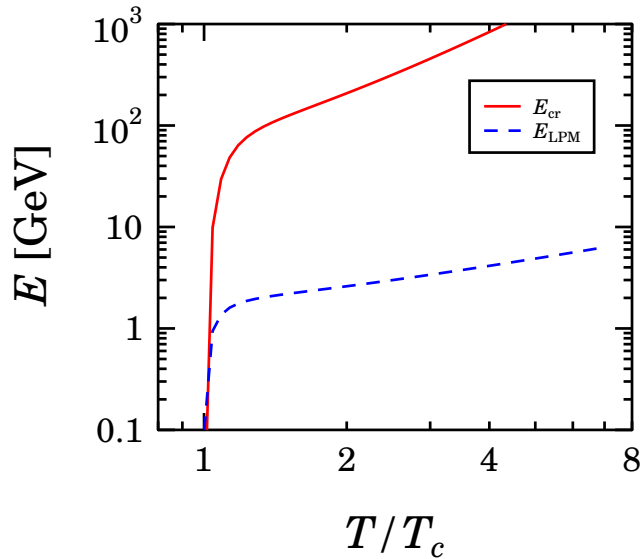}}
 \caption{The temperature dependence of the energy scales $E_{\rm cr}$ and 
    $E_{\rm LPM}$ for the radiative energy loss. The length of the medium
    was set to $L=5\,$fm, and $T_c=170\,$MeV.
  \label{fig: Ecr}}
\end{figure}

\begin{figure}[ht]
 \centerline{\includegraphics[scale=0.9]{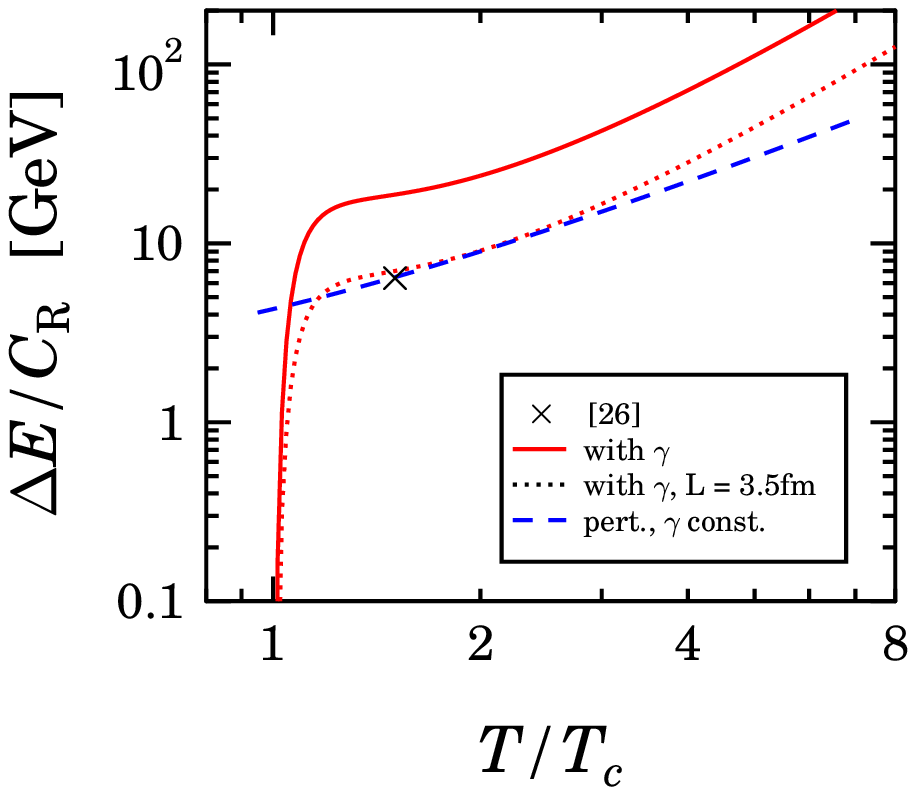}}
 \caption{The radiative energy loss from Eq.~(\ref{eq: Delta E}) for
    $L=5\,$fm, unless indicated otherwise.
    The symbol is an estimate from Ref.~\cite{BDMPS}.
    The dashed line results from a 2-loop running coupling, with 
    $m_{\rm D}$ set to $m_{\rm D,0}$ and a constant $\gamma$ to match the
    result from \cite{BDMPS}.
    The full line is the estimate with the $T$-dependent width and the 
    adjusted coupling, as is the dotted line where $L=3.5\,$fm was assumed 
    to illustrate the scaling behavior.
  \label{fig: DeltaE}}
\end{figure}

There are several interesting implications of the characteristic temperature dependence of the damping rate besides those for the screening properties discussed above. As the inverse of the mean free path $\lambda$, it is closely related to transport properties as, e.\,g., equilibration times. Expecting a critical slowing down, this link provides another indication that the width has to become small near $T_c$.
Another quantity, which is of particular interest with regard to the interpretation of SPS and RHIC experiments, is the radiative energy loss of hard quarks and gluons transversing the plasma.
The results derived in \cite{BDMPS} for a system of length $L$ under the assumption of (several) independent scatterings, i.\,e.\ $m_{\rm D} \gg \gamma$, which is not unrealistic in the situation of interest, see below, are characterized by the energy scale
\be
  E_{\rm cr} = \gamma m_{\rm D}^2 L^2 \, .
\ee
In the Landau-Pomeranchuk-Migdal regime, for parton energies $E > E_{\rm cr}$, the energy loss reads \cite{BDMPS}
\be
  -\Delta E 
  = 
  \T\frac18\, C_R\, \alpha \gamma m_{\rm D}^2 L^2 \ln \gamma L \, ,
  \label{eq: Delta E}
\ee
where $R$ indicates the color representation of the parton.
As argued in \cite{DumiP}, the expected `critical' behavior of the screening mass would lead to a reduced energy loss at temperatures near $T_c$, possibly explaining the absence of jet quenching at SPS energies
\cite{Wang}.
With the temperature dependence of the width suggested here, this effect would be even more pronounced. At the same time, the adjustment of the parameters to lattice data may allow for a realistic estimate.
In order to describe the Debye mass in a simple way, without any assumptions, I make use of the empirical observation that for the relevant temperatures
\be
  m_{\rm D} \approx 2.7 \gamma \, ,
\ee
cf.\ Fig.~\ref{fig: m_debye}.
The resulting behavior of the energy scale $E_{\rm cr}$, shown in Fig.~\ref{fig: Ecr}, 
changes drastically at 
\be
  T_E \approx 1.3T_c \, .
\ee
Close to $T_c$, the energy $E_{\rm cr}$ becomes very small. Already for slightly larger temperatures it is well above the scale $E_{\rm LPM} = m_{\rm D}^2/\gamma$ which is relevant for the energy loss in the Bethe-Heitler regime.
A similar sudden onset, also at $T_E$, is found for the medium induced energy loss, cf.\ Fig.\ref{fig: DeltaE}.
For larger temperatures the estimate agrees basically with other results, while close to $T_c$ it becomes very small and would be hard to observe experimentally.
Similar small results have been obtained for the corresponding SPS energies in Ref.~\cite{GyulLV}.

In summary, it has been argued from the reduced number of degrees of freedom near the transition temperature that the width of hard excitations has to become small near $T_c$.
While for QCD this was demonstrated under the assumption of a Lorentz spectral function, the propagator may have a more complicated pole structure.
For hot QED (where no magnetic mass exists), the fermion propagator has been calculated in an Bloch-Nordsieck approach \cite{IancB}. The result is infrared finite, and an {\em entire} function of the energy. The spectral function is nonetheless strongly peaked, with a characteristic width $\sim e^2 \ln(1/e)T$. Since in Fourier space it decreases faster than an exponential, the effect for the entropy will be even more pronounced than for a Lorentzian.
While the situation may be different in QCD, a spectral function with $\rho(t) \sim t^{-1}$, which has little effect on the entropy, seems hard to explain.
Although from thermal properties obtained in lattice QCD little can be inferred about the analytic structure of the propagator, the general result of small widths near $T_c$ is arguably robust.
This implies a characteristic change of several phenomenologically relevant quantities at $T^\star \approx T_\gamma \approx T_E \approx 1.3 T_c$.

\vskip 5mm
\noindent
{\bf Acknowledgements:}
I acknowledge stimulating discussions on this and related subjects with W.~Cassing, A.~Dumitru, F.~Gelis, S.~Leupold, C.~Lorenz, U.~Mosel, and R.~Pisarski. This work is supported by BMBF.

\appendix

\section{Some properties of \bm{\Delta}{\footnotesize\bm{s}} 
    \label{sec: appendix A}}

In the following it is argued that the entropy is generally increased for a non-zero width, i.\,e.\ $\Delta s > 0$. Furthermore, considering the Lorentz spectral function, the expansion of $s_L(m,\gamma)$ is calculated for small arguments. Finally it is proved that $s_L(m,\gamma)$ is for $m = \gamma$ equal to the Stefan-Boltzmann value $s_0 = s^{(0)}(m=0) = \frac{4\pi^2}{90}\,T^3$.
\\[3mm]
In order to prove (under the assumption of a unique dispersion relation $\omega_k$) that $\Delta s > 0$, consider the relevant integral in (\ref{eq: Delta s}) in the form
\[
  I
  = 
  \int_0^\infty d\omega\, 
    \frac{\p n(\omega)}{\p T}\, f(\lambda(\omega)) \, ,
\]
where $f = {\rm arctan}\lambda-\lambda/(1+\lambda^2)$, and $\lambda = \Im\Delta/\Re\Delta$. Changing the integration variable to $\lambda$, 
\[
  I 
  = 
  \int_{-\infty}^\infty d\lambda\,
    \frac{\p \omega}{\p \lambda}\, \frac{\p n(\omega(\lambda))}{\p T}\, 
    f(\lambda) \, ,
\]
the integrand becomes a product of three factors, of which $f(\lambda)$ is an odd function. The sign of $I$ is, thus, determined only by the other two terms, which can be discussed on the basis of the relation $\omega(\lambda)$, whose inverse is shown in Fig.~\ref{fig: intgrd_Delta_s}.
\begin{figure}[ht]
 \hskip -1em
 \includegraphics[scale=0.75]{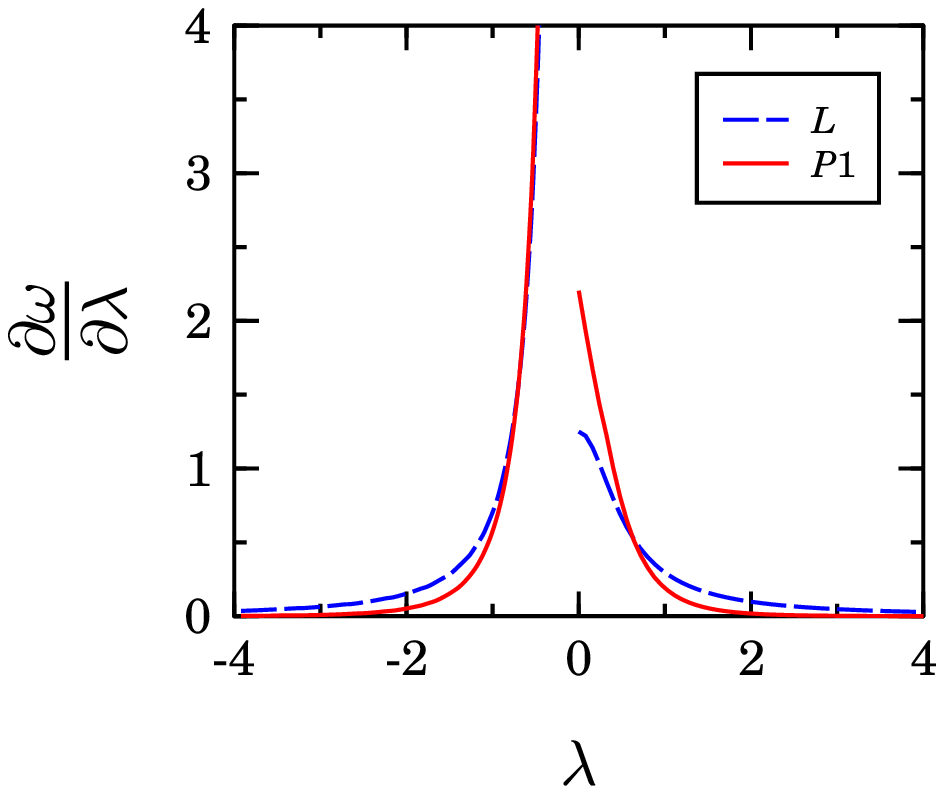}
 \hfill
 \includegraphics[scale=0.75]{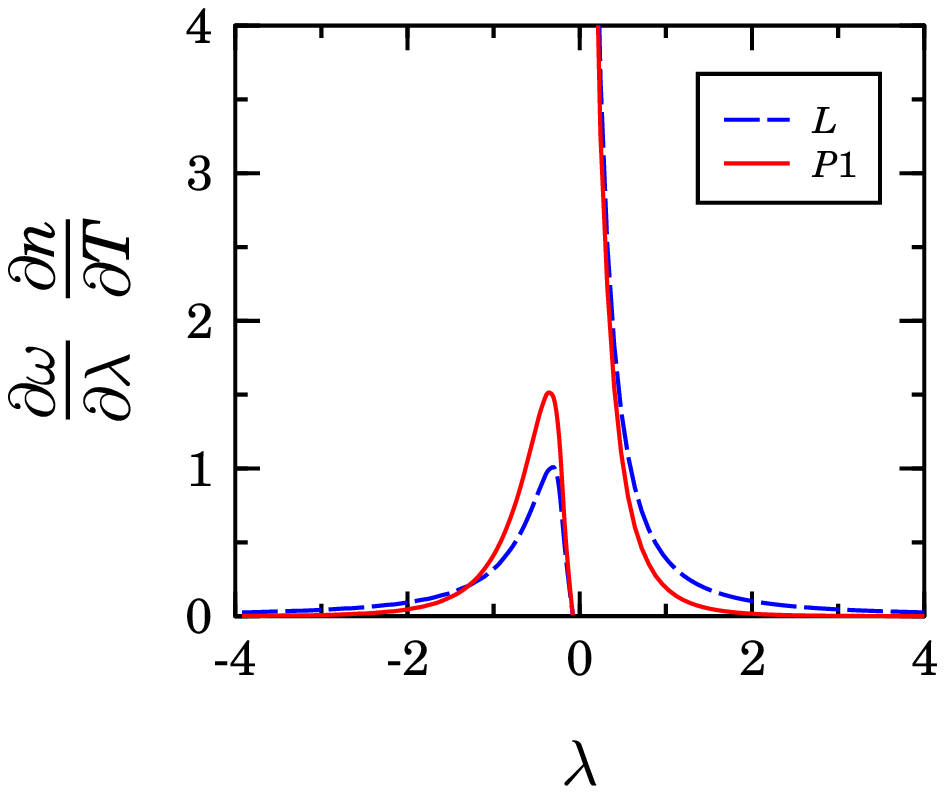}
 \caption{Two functions used in the argumentation that $\Delta s > 0$, 
    for the case of the propagators from Fig.~\ref{fig: types of 
    prop}.
  \label{fig: appx}}
\end{figure}
In particular, $\omega \rightarrow \{ 0, \omega_k-0, \omega_k+0, \infty \}$ corresponds to $\lambda \rightarrow \{ 0^+, +\infty, -\infty, 0^- \}$.
Then, via the inverse derivative $\p\lambda/\p\omega$ and the properties of the propagator listed in Section 2, it is easily inferred that $\p\omega/\p\lambda$ is positive and that it vanishes for $|\lambda| \rightarrow \infty$, is non-zero for $\lambda \rightarrow 0^+$, and diverges for $\lambda \rightarrow 0^-$, cf.\ Fig.~\ref{fig: appx}. The product $\p\omega/\p\lambda\, \p n/\p T$, however, vanishes for small negative $\lambda$ due to the second factor, which for the corresponding large $\omega$ is exponentially suppressed. At small positive $\lambda$, on the other hand, this factor is Bose enhanced, $\p n/\p T \sim \omega^{-1}$, while it goes to the value $\p n/\p T|_{\omega_k}$ for $\lambda \rightarrow\pm\infty$.
Therefore, the integral of the product $(\p\omega/\p\lambda)\, (\p n/\p T)$ over $[-\infty,0]$ is finite, while for the interval $[\varepsilon,+\infty]$ it approaches $+\infty$ for $\varepsilon \rightarrow 0$. This shows that the integral $I$, whose integrand includes the odd function $f(\lambda)$, is positive (and finite due to $f \sim \lambda^3$ for small $\lambda$), hence $\Delta s > 0$.
\\[3mm]
Turning now to the expansion of the entropy $s_L(m,\gamma)=s^{(0)}(m)+\Delta s_L(m,\gamma)$ for a Lorentzian spectral function, I introduce the notation $\Gamma=2\gamma$ and follow the remark below Eq.~(\ref{eq: Delta s}), writing
\[
  \Delta s_L(m,\Gamma)
  =
  \int_0^\infty\frac{d\omega}\pi\, \frac{\p n}{\p T}
  \int_{k^3} \l( h - \Gamma \frac{\p h}{\p \Gamma}\, \r) ,
  \label{eq: structure s}
\]
where $h = {\rm arctan}(\Gamma\omega/(\omega_m^2-\omega^2))$.
Considering first the case $m=0$, i.\,e., $\omega_m = k$, the two terms of the $k$-integrand decrease as $k^{-2}$.
In the subtracted integral
\[
  {\cal I}
  =
  \int_{k^3}\,
    \l( \frac{\p h}{\p \Gamma} - \frac{\omega}{k^2} \r)
  =
  \frac\omega{2\pi^2}
  \int_0^\infty dk\, k^2 \l(
   \frac{k^2-\omega^2}{(k^2-\omega^2)^2+\Gamma^2\omega^2}-\frac1{k^2}
  \r) ,
\]
the substitutions $k = x\omega$ and $a = \Gamma/\omega$ lead to
\bean
  {\cal I}
  &=&
  \frac{\omega^2}{2\pi^2}\int_0^\infty dx\, x^2 \l(
   \frac{x^2-1}{(x^2-1)^2+a^2}-\frac1{x^2}
  \r)
  \\
  &=&
  -\frac{\omega^2}{2\pi^2}\, \frac\pi{2\sqrt2}\, \l( \sqrt{1+a^2}-1 \r)^{1/2}
  =
  -\frac{\omega\Gamma}{8\pi}+ \ldots
\eean
The remaining $\omega$-integral yields
\[
  {\cal J}
  =
  \int_0^\infty\frac{d\omega}\pi\, \frac{\p n}{\p T}\, {\cal I}
  =
  -\frac{T}{24}\, \Gamma + \ldots
\]
From $\Delta s_L(m, \Gamma) = \int\!d\Gamma\, {\cal J} - \Gamma {\cal J}$, and since $\Delta s_L(m, \Gamma=0) = 0$ (for any $m$), one obtains
\[
  \Delta s_L(m=0,\Gamma)
  =
  \frac{T}{48}\, \Gamma^2 + \ldots
\]
Furthermore, it is obvious that derivatives of any order of $\Delta s_L(m,\Gamma)$ with respect to $m$ vanish at $\Gamma=0$.
Therefore, the leading term in $m$ in the expansion of the total entropy comes entirely from the contribution $s^{(0)}(m)$, which is well known, and one arrives at the expression (\ref{eq: sL expansion}).
\\[3mm]
Finally, the fact that $s_L(m,\gamma=m) = s_0$ holds not only in the limit of small $m$ and $\gamma$ is readily proven by verifying $\p s_L(m,\gamma=m) / \p m = 0$. After taking the derivative of the integrand of the total entropy \footnote{
    Note that the integrand of $s_L$, contrary to that of $\Delta s_L$,
    is a smooth function of $m$.},
the $k$-integration yields zero, indeed.
As an aside it is mentioned that for $\gamma \ge m$, the poles of the propagator (\ref{eq: Delta_L}) are purely imaginary for some range of momenta, see footnote~[34]. %\ref{footnote1}.

\end{document}